\documentclass[pdflatex,sn-vancouver-num]{sn-jnl}

\usepackage{graphicx}
\usepackage{multirow}
\usepackage{amsmath,amssymb,amsfonts,bm}
\usepackage{amsthm}
\usepackage{mathrsfs}
\usepackage[title]{appendix}
\usepackage{xcolor}
\usepackage{textcomp}
\usepackage{manyfoot}
\usepackage{booktabs}
\usepackage{longtable}
\usepackage{array}
\usepackage{siunitx}
\usepackage{tabularx}
\usepackage{adjustbox}
\usepackage{makecell}
\usepackage{ragged2e}
\usepackage[caption=false]{subfig}

\begin{document}

\title{A quantum-like benchmark for context-sensitive associative memory with adaptive plasticity}

\author*[1]{\fnm{Yashine H.} \sur{Goolam Hossen}}\email{yhgoolam@uwaterloo.ca}

\author[2]{\fnm{Lea} \sur{Gassab}}\email{lgassab@uwaterloo.ca}

\author[1]{\fnm{Travis J. A.} \sur{Craddock}}\email{travis.craddock@uwaterloo.ca}

\affil*[1]{\orgdiv{Departments of Biology and Physics \& Astronomy, Waterloo Institute for Nanotechnology}, \orgname{University of Waterloo}, \orgaddress{\city{Waterloo}, \state{Ontario}, \country{Canada}}}

\affil[2]{\orgdiv{Departments of Biology, Chemistry, Physics \& Astronomy, Waterloo Institute for Nanotechnology}, \orgname{University of Waterloo}, \orgaddress{\city{Waterloo}, \state{Ontario}, \country{Canada}}}

\abstract{Learning and memory require a balance between plasticity and stability: synaptic connections must encode new information without collapsing, saturating, or erasing previously useful structure. Associative-memory models can appear to learn successfully when fixed background connectivity already carries part of the task, making it difficult to distinguish genuine recall dynamics from structural assistance. We test this issue using an order-sensitive adaptive-plasticity benchmark for staged associative recall. The benchmark compares a quantum-like associative-memory model with matched real-valued no-phase and Markov-rate controls under the same task schedule, perturbation profiles, weak-support conditions, and plasticity settings. Here, ``quantum-like'' refers to the modeling formalism, not to a biological claim about quantum computation. We first screen weak structural support and then fix a conservative operating point for factorial comparisons across model families and plasticity mechanisms. The useful weak-support regime is narrow and non-monotonic. Weak structure alone does not rescue recall in the no-plasticity ablation, whereas most useful recall gains arise from adaptive plasticity, especially homeostatic stabilization. The Markov-rate control often achieves stronger raw recall, but the quantum-like model more consistently preserves order sensitivity and stage-dependent organization. These results do not support a universal quantum-like advantage. Instead, they show that model classes are better distinguished by a multi-objective profile combining recall, temporal organization, and context sensitivity than by any single recall score. The benchmark therefore provides a controlled framework for studying context-sensitive memory dynamics under weak support, regulated plasticity, and matched classical comparison.}


\maketitle

\tableofcontents

\section{Introduction}\label{sec:introduction}

Learning and memory face a fundamental tension in both biology and modeling: synapses must remain plastic enough to encode new associations, yet stable enough to avoid runaway excitation, collapse, or erasure of previously useful structure. This tension is already implicit in Hebb's original proposal for activity-dependent strengthening of synaptic connections between co-active neurons \cite{hebb1949organization}, in the experimental establishment of long-term potentiation as a durable substrate of memory-related change \cite{blisslomo1973}, and in spike-timing-dependent synaptic modification as a temporally structured form of activity-dependent plasticity \cite{bi1998}. Subsequent work on associative memory showed that distributed recall can emerge from collective interactions among many neurons or units coupled through recurrent synaptic weights \cite{hopfield1982,dayanabbott2001}, but it also made clear that unconstrained plasticity can rapidly destabilize such models or require normalization or stabilizing constraints to remain interpretable \cite{oja1982}. Homeostatic plasticity regulates synaptic gain around functional set-points \cite{turrigiano2012}; heterosynaptic plasticity redistributes weight beyond directly active synapses and supports competition \cite{chistiakova2015,jenks2021}; and structural plasticity enables the longer-timescale formation, stabilization, or elimination of synaptic contacts \cite{holtmaat2009}. Any useful associative-memory model therefore has to answer not only whether a learned pattern can be recovered, but how plasticity, stability, and circuit structure are balanced while recovery remains interpretable.

A second challenge is evaluative. In a staged memory task, what should count as success? Classical attractor models are often judged by endpoint recovery, yet many memory phenomena depend on temporal context, sequence, and path dependence rather than on static completion alone. More broadly, dynamical-systems perspectives emphasize that adaptive neural behavior depends on trajectories through state space and on transitions between regimes, not only on fixed-point outcomes \cite{breakspear2017}. In the present task, the ordering of past association, rest, novel exposure, rest, and cue-triggered re-exposure is therefore part of the phenomenon itself. A model that reaches high recall only by weakening the intended stage structure is not solving the same problem as one that preserves selective stage structure and order dependence.

These considerations motivate testing a quantum-like formulation rather than assuming that endpoint recall alone is the relevant target. Here, ``quantum-like'' refers to a complex state-space formalism with interference-capable and order-sensitive dynamics; it does not imply microscopic quantum computation in neural tissue. In cognitive modeling, quantum-probabilistic formalisms have been used to capture context dependence, interference, and order effects that are awkward to express in classical probabilistic frameworks \cite{busemeyerbruza2012,pothos2013,wang2014}. Recent work by Scholes and colleagues has further shown that related quantum-like state-space structure can emerge from nonlinear classical synchronization networks, producing superposition-like states, interference-like behavior, quantum-like information encoding, and context-encoded dynamics without invoking microscopic quantum computation in neural tissue \cite{scholes2024qlstates,scholes2026dynamics,amati2025qlbits,amati2025encoding}. That literature does not by itself establish that a quantum-like model should outperform classical alternatives in the present setting. It does, however, motivate a benchmark in which order sensitivity and stage structure are treated as explicit targets alongside recall.

The present study therefore has both a mechanistic aim and a benchmarking aim. Mechanistically, we ask which ingredients actually carry recall once weak structural support, adaptive plasticity, and matched classical controls are placed in the same staged task. From the benchmarking side, we ask how to avoid three common interpretive failures: mistaking fixed background support for learned recall, allowing structural support to become an uncontrolled source of dense connectivity, and improving recall only at the cost of the task's intended stage structure. More generally, a benchmark is only informative if it resists shortcut solutions, keeps the target capability aligned with the evaluation objective, and reduces the risk that apparent gains are driven mainly by overfitting to the benchmark itself \cite{Geirhos2020Shortcut,Lever2016ModelSelection,Weber2019Benchmarking,Bowles2024SubtleArt}.

To keep these issues separate, we evaluate every model along three predefined summary families:
recall, stage structure, and post-learning order asymmetry. These quantities track recovery magnitude, preservation of the intended stage sequence, and path dependence, respectively. Support selection, control calibration, and model comparison are all disciplined by the same benchmark rules, so the reported comparisons are not chosen retrospectively to optimize a single scalar score. 

To address these questions, we use a two-stage design. First, we sweep the strength of a sparse structural floor and retain an operating point only when the full quantum-like model improves while the no-plasticity quantum-like ablation remains unchanged on the screening metrics. Second, we freeze that operating point and perform a factorial comparison across perturbation profile, model class, support condition, and plasticity combination. The paper is therefore organized around three questions: which support regime remains useful once trivial scaffold rescue is excluded, which plasticity mechanisms actually carry the recall effect, and whether a quantum-like formulation offers benchmark-relevant advantages once real-valued no-phase and Markov-rate baselines are matched on the same sequentially structured task. 

The remainder of this paper is organized as follows. Section~\ref{sec:methods} defines the staged recall task, model families, weak-support construction, plasticity mechanisms, and evaluation summaries. Section~\ref{sec:results} presents the results in the benchmark sequence: weak-support selection, profile-resolved dynamic checks, factorial plasticity decomposition, and matched model-family comparison. Section~\ref{sec:discussion} interprets these findings in the broader context of adaptive memory modeling and restrained quantum-like comparison. Section~\ref{sec:conclusion} summarizes the main conclusions and outlines future directions. Additional implementation details, tables, and diagnostic figures are provided in the \hyperref[sec:supplementary]{Supplementary Information}.

\section{Methods}\label{sec:methods}

\begin{figure*}[t]
\centering
\includegraphics[width=1\linewidth]{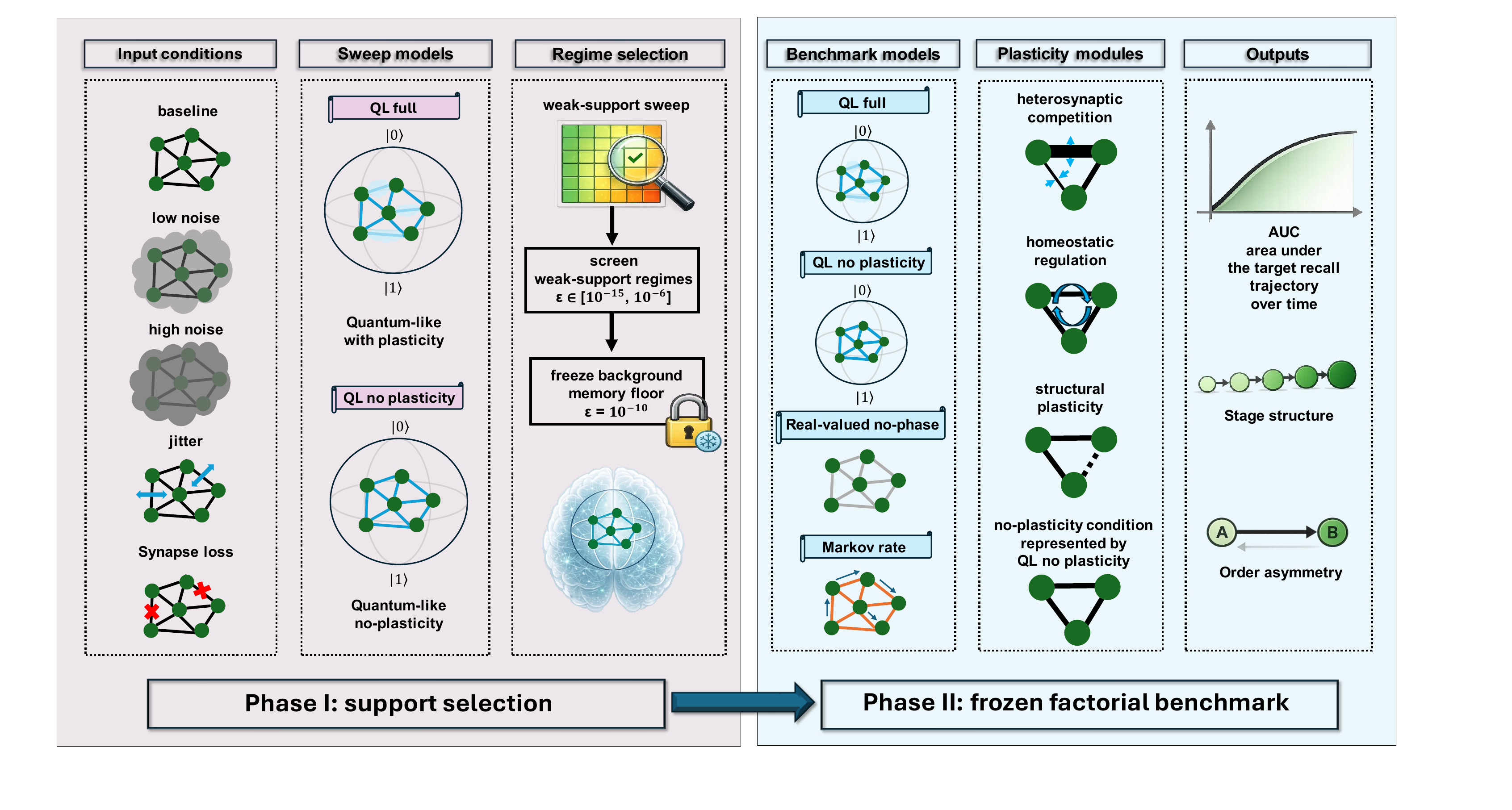}
\caption{Schematic overview of the executed two-phase benchmark workflow. In Phase I (support selection), the same staged memory task is evaluated under five perturbation profiles (baseline, low noise, high noise, jitter, and synapse loss), and a weak-support sweep is performed only for the full quantum-like model and its no-plasticity quantum-like ablation. The sweep includes the zero-floor reference together with nonzero weak-support values from \(10^{-15}\) to \(10^{-6}\). A conservative nonzero operating point is then fixed by freezing the background memory floor at \(\epsilon=10^{-10}\), preserving a minimal positive scaffold rather than removing support entirely. In Phase II (frozen factorial benchmark), this selected operating point is held fixed while the benchmark compares the full quantum-like model, the no-plasticity quantum-like ablation, the real-valued no-phase control, and the Markov-rate control. Heterosynaptic competition, homeostatic regulation, and structural plasticity define the adaptive ingredients of the factorial benchmark. Final outputs summarize recall-stage area under the curve, stage structure, and order asymmetry.}
\label{fig:workflow_overview}
\end{figure*}

Figure~\ref{fig:workflow_overview} summarizes the two-stage benchmark design. We first selected a weak-support operating point, then froze that setting for factorial plasticity and model-family comparisons. The three primary summaries are recall AUC, stage structure, and order asymmetry. Recall AUC denotes the time-integrated \(A\)-channel trajectory during the final A-cued recall stage; it is not a classification-style ROC AUC. Larger values indicate stronger or more sustained recovery of the cued pattern. Stage structure denotes preserved alignment with the intended task logic: the past stage should favor the learned \(A/B\) channels, the novel-exposure stage should favor \(C\), the final A-cued recall stage should favor \(A\), and the intervening rest stages should remain comparatively quiet. Order asymmetry measures whether short \(A\!\rightarrow\!C\) and \(C\!\rightarrow\!A\) probe sequences produce different readouts from the same post-learning state. Table~\ref{tab:reader_guide} provides an interpretive guide to these summaries. Spectral and modular diagnostics are used only to interpret the dynamics behind them. Additional workflow, calibration, simulator, and aggregation
details are provided in the \hyperref[sec:supp_methods]{Supplementary Methods} and \hyperref[sec:supp_math_details]{Supplementary mathematical details}, with extended quantitative tables in \hyperref[sec:supp_tables]{Supplementary tables}.

\begin{table}[t]
\centering
\small
\caption{Operational interpretation of the primary benchmark summaries.}
\label{tab:reader_guide}
\renewcommand{\arraystretch}{1.50}
\setlength{\tabcolsep}{3pt}
\begin{tabular}{
    p{0.15\linewidth}
    @{\hspace{0.02\linewidth}}
    p{0.25\linewidth}
    @{\hspace{0.02\linewidth}}
    p{0.19\linewidth}
    @{\hspace{0.02\linewidth}}
    p{0.19\linewidth}
}
\toprule
Summary & What it measures & Higher / more positive values indicate & Lower / more negative values indicate \\
\midrule
Recall-stage AUC & Time-integrated target-channel recall during the final A-cued stage. & The correct A-channel stays elevated for longer, or at a stronger sustained level, during final recall. & Recall is weak, brief, unstable, or displaced into the wrong channels. \\
Stage-structure score & How well activity stays aligned with the intended stage logic: past favors A/B, the new stage favors C, and the final stage favors A while rest remains quiet. & The model preserves the intended temporal organization of the task. & The task stages blur together, off-target channels dominate, or rest activity becomes too strong. \\
Order asymmetry & Difference between short A$\rightarrow$C and C$\rightarrow$A probe sequences applied from the same post-learning state. & The system is more sensitive to sequence and context. & The system behaves more similarly for both probe orders and is less sequence-sensitive. \\
Paired $d_z$ & Standardized within-condition effect size for matched contrasts. & The contrast is consistently positive across paired seeds. & The contrast is small, inconsistent, or favors the comparison condition. \\
\bottomrule
\end{tabular}
\end{table}

\subsection{Task design and state representations}
The benchmark task was designed to test cue-based associative memory in a staged setting rather than memory retrieval in a single undifferentiated epoch. The network contained $N=90$ nodes divided into three fixed stimulus-defined groups,
\begin{equation}
A=\{0,\ldots,24\},\qquad B=\{25,\ldots,49\},\qquad C=\{50,\ldots,74\},
\end{equation}
with the remaining nodes serving as unstimulated background degrees of freedom. The temporal protocol consisted of five ordered epochs,
\begin{equation}
\mathrm{past}(A/B) \rightarrow \mathrm{rest}_1 \rightarrow \mathrm{new}(C) \rightarrow \mathrm{rest}_2 \rightarrow \mathrm{recallA}(A),
\end{equation}
with durations $T_{\mathrm{past}}=3500$, $T_{\mathrm{rest1}}=1200$, $T_{\mathrm{new}}=2000$, $T_{\mathrm{rest2}}=800$, and $T_{\mathrm{recallA}}=1500$ integration steps at $\Delta t=0.03$. During the past epoch, templates $A$ and $B$ alternated every $200$ logged steps; during the new epoch, template $C$ was applied; and during the final recall epoch, template $A$ was re-presented. Stimulus injection used a larger onset mixture coefficient and a smaller sustained coefficient,
\begin{equation}
\alpha_{\mathrm{switch}}=0.25,\qquad \alpha_{\mathrm{cont}}=0.04,
\end{equation}
which were subsequently rescaled for the classical controls by a fitted stimulus-amplitude scale factor. Additional rationale for the cue-onset and sustained-cue amplitudes, including their operational rather than biological interpretation, is given in Supplementary Section~\ref{sec:supp_operational_settings} and Table~\ref{tab:param_rationale}.

The five stages separate association formation, rest-state suppression, new-input encoding, and final cued recovery after an intervening history. The task therefore tests not only whether recall occurs, but whether it occurs without collapsing the staged temporal structure.

For the quantum-like model, the instantaneous state was a normalized complex vector
\begin{equation}
\psi(t) \in \mathbb{C}^{N}, \qquad \lVert \psi(t) \rVert_2 = 1,
\end{equation}
whereas the real control used a normalized real vector $x(t)\in\mathbb{R}^N$ and the continuous-time Markov-rate control used a normalized nonnegative vector $p(t)\in\mathbb{R}_{\ge 0}^N$ with $\sum_i p_i=1$. All runs were initialized from seed-specific random normalized states, with the learned fast and slow connectivity matrices initialized at zero; the exact initialization convention is given in Supplementary Section~\ref{sec:supp_initialization}.

Stimulus templates were defined by leakage-smoothed indicator vectors and injected with a larger amplitude at switching times and a smaller amplitude during sustained presentation,
\begin{equation}
\psi \leftarrow \frac{(1-\alpha)\psi + \alpha S}{\lVert (1-\alpha)\psi + \alpha S \rVert_2},
\end{equation}
and analogously for the real and probabilistic controls. Here \(S\in\{S_A,S_B,S_C\}\) denotes the currently active stimulus template. The templates \(S_A\), \(S_B\), and \(S_C\) are the leakage-smoothed indicator vectors supported on the node groups \(A\), \(B\), and \(C\), respectively. Operationally, the same staged memory task was therefore applied to three state formalisms: a complex quantum-like state, a real-valued no-phase state, and a probability-like Markov-rate state.

\subsection{Quantum-like and classical dynamics}
All model families were evaluated under the same task schedule, stage boundaries, and adaptive-operator logic. The comparison therefore tests how each state formalism behaves under matched perturbations. 

Let $A_{\mathrm{eff}}(t)$ denote the instantaneous effective connectivity operator obtained from the learned fast and slow matrices, together with the weak-support floor when support is enabled. Its explicit construction is given in Eq.~(\ref{eq:backbone}). The factor $s_{\mathrm{dyn}}$ is a global dynamical-rate scale: for the quantum-like model it is kept at the native value 1, whereas for the classical controls the corresponding rate scale is calibrated within each matched condition on the calibration split. The quantum-like state evolved under a Hamiltonian-like effective connectivity operator,
\begin{equation}
H_{\mathrm{eff}}(t) = s_{\mathrm{dyn}}\,A_{\mathrm{eff}}(t),
\end{equation}
according to
\begin{equation}
\frac{d\psi}{dt} = -iH_{\mathrm{eff}}(t)\psi,
\label{eq:ql_dynamics}
\end{equation}
which was integrated numerically with a fourth-order Runge-Kutta (RK4) scheme. State noise, when present, was added as complex Gaussian noise with the same $\sqrt{\Delta t}$ scaling used for the real control. The real-valued control followed the same task and adaptive operator but used the real-valued dynamics without a phase degree of freedom,
\begin{equation}
\frac{dx}{dt}=s_{\mathrm{dyn}}A_{\mathrm{eff}}(t)x,
\end{equation}
with renormalization after each step. The Markov-rate control used a probability-conserving continuous-time generator built from the same instantaneous effective matrix. Writing
\begin{equation}
W_{ij}(t)=\begin{cases}
|A_{\mathrm{eff},ij}(t)|, & i\neq j,\\
0, & i=j,
\end{cases}
\qquad
D_{ii}(t)=\sum_j W_{ij}(t),
\end{equation}
and $L(t)=D(t)-W(t)$, the Markov-rate state followed
\begin{equation}
\frac{dp}{dt}=-s_{\mathrm{dyn}}L(t)p,
\end{equation}
with normalization and nonnegativity enforcement after each step. Because \(A_{\mathrm{eff}}\) is symmetrized before constructing \(W\), the resulting \(W\) is symmetric, so the row and column sums of \(L\) vanish consistently and the continuous-time update is probability conserving before numerical renormalization. The role of these controls was not merely to provide lower-performing baselines, but to test whether high recall could be obtained at the cost of losing temporal order sensitivity and stage selectivity. 

\subsection{Adaptive operator and plasticity terms}

The effective operator combined a fast adaptive component, a slower consolidating component, and optional additional plasticity rules. At each time point, the instantaneous Hebbian drive was defined as
\begin{equation}
H_{ij}(t)=\mathrm{Re}\!\left[\psi_i(t)\,\overline{\psi_j(t)}\right]M_{ij},
\end{equation}
for the quantum-like model, with analogous real-valued definitions for the controls. Here \(M\) is a module mask inferred from \(|A_{\mathrm{slow}}|\) by normalized spectral clustering \cite{ng2001spectral,newman2006}; the explicit construction is given in Supplementary Section~\ref{sec:supp_module_mask}. The fast component followed
\begin{equation}
A_{\mathrm{fast}} \leftarrow (1-\gamma_{\mathrm{fast}})A_{\mathrm{fast}} + \eta_{\mathrm{fast}}H,
\end{equation}
with \(\gamma_{\mathrm{fast}}=2\times10^{-3}\) and \(\eta_{\mathrm{fast}}=0.025\), and the slow component followed
\begin{equation}
A_{\mathrm{slow}} \leftarrow (1-\gamma_{\mathrm{slow}})A_{\mathrm{slow}} + \kappa A_{\mathrm{fast}},
\end{equation}
with \(\gamma_{\mathrm{slow}}=2\times10^{-5}\) and \(\kappa=3\times10^{-4}\). These fixed operational parameters separate rapid task-driven adaptation from slower consolidation.

Three additional plasticity mechanisms were toggled factorially: homeostatic stabilization, heterosynaptic competition, and structural plasticity. Homeostatic plasticity acted as a stabilizing process that limited excessive slow-matrix load. Heterosynaptic plasticity penalized weakly matched fast edges in proportion to node-level Hebbian load, implementing competitive redistribution rather than a second reinforcement rule. Structural plasticity maintained a slowly evolving trace of Hebbian drive and used it to create, retain, or prune weak slow edges. The full factorial design therefore comprised the \(2^3\) combinations of homeostatic, heterosynaptic, and structural plasticity. In this factorial grid, the ``none'' condition meant that the three additional mechanisms were switched off. It did not denote the no-plasticity quantum-like ablation. The full quantum-like model retained the base fast/slow adaptive update unless it was explicitly labeled as the no-plasticity ablation.

In conceptual terms, the three mechanisms represent distinct forms of adaptation with different functional roles. Homeostatic plasticity maintains synaptic load within a functional range and limits runaway strengthening or collapse. Heterosynaptic plasticity redistributes synaptic influence beyond the directly reinforced synapses. Structural plasticity modifies the effective support pattern by creating, pruning, or consolidating connections. The detailed executed update rules and parameter values are provided in Supplementary Section~\ref{sec:supp_plasticity_rules}, with operational parameter rationale summarized in Table~\ref{tab:param_rationale}.

\subsection{Weak structural support}
The weak-support factor was introduced to test whether a small persistent scaffold could stabilize recall without replacing the learned dynamics. The scaffold was implemented as a floor-only rule on the support of the learned matrix, not as an unrestricted additive network. First,
\begin{equation}
A_{\mathrm{plastic}}^{\star}=\mathrm{clip}\!\left(\mathrm{sym}_0\big(A_{\mathrm{slow}}+A_{\mathrm{fast}}\big)\right),
\end{equation}
where \(\mathrm{sym}_0(X)=\tfrac12(X+X^\top)-\mathrm{diag}(X)\) enforced symmetry with zero diagonal, and \(\mathrm{clip}\) denoted elementwise clipping to the fixed simulator bound. When support was enabled, a sparse modular floor matrix $B_{\mathrm{floor}}$ was constructed on the nonzero support of $A_{\mathrm{plastic}}^{\star}$ and rescaled row-wise to meet a row-$L^1$ budget $\epsilon$. The effective matrix was then
\begin{equation}
A_{\mathrm{eff},ij}=\operatorname{sign}\!\left(A_{\mathrm{plastic},ij}^{\star}\right)
\max\!\left(|A_{\mathrm{plastic},ij}^{\star}|, |B_{\mathrm{floor},ij}|\right).
\label{eq:backbone}
\end{equation}
Thus, the scaffold can stabilize existing support without creating new dense connectivity. We then swept \(\epsilon\) and selected the operating point before interpreting the downstream factorial results. Candidate floors were accepted only if they improved the full quantum-like model without materially improving the no-plasticity ablation on recall AUC, support density, or zero-edge fraction. The selected value was \(\epsilon=10^{-10}\) under the sparse modular topology; the exact sweep logic is described in Supplementary Section~\ref{sec:supp_weak_support}, and the exact sweep values are reported in Table~\ref{tab:sweep_exact}.

\subsection{Benchmark aggregation and summary flow}

The simulation workflow separated support selection, control calibration, and frozen-condition evaluation. Backbone-on denotes the selected weak-support condition with \(\epsilon=10^{-10}\), whereas backbone-off denotes the zero-floor reference \(\epsilon=0\). Individual trajectories were segmented by task stage and then aggregated into the predefined recall AUC, stage-structure, and order-asymmetry summaries. Model, backbone, and plasticity contrasts were computed only after the support level and calibration settings were frozen, so the reported comparisons reflect matched evaluation conditions rather than retrospective optimization of a single scalar score. All primary contrasts were computed on the disjoint evaluation split, using seeds 11--40 (\(n=30\) paired evaluation seeds per matched condition). Calibration seeds were used only to tune the classical-control scale factors and were excluded from the reported model comparisons. For the principal uncertainty summaries, matched differences were first averaged within each evaluation seed across the relevant matched condition cells, and the resulting seed-level difference vector was resampled with replacement. 
Seed-level \(95\%\) confidence intervals were estimated by percentile bootstrap resampling of the \(30\) evaluation seeds, using \(20{,}000\) resamples. Positive-fraction summaries in Fig.~\ref{fig:matched_positive_fraction} were computed over matched evaluation cells and are reported descriptively. The detailed stage-aggregation formulas are provided in Supplementary Section~\ref{sec:supp_pipeline_aggregation}, the spectral and modular diagnostics and support summaries in Supplementary Section~\ref{sec:supp_spectral_diagnostics}, and the paired-contrast formulas in Supplementary Section~\ref{sec:supp_paired_contrasts}.

\subsection{Perturbation profiles, calibration, and metrics}
Five perturbation profiles were used: baseline, low noise, high noise, jitter, and synapse loss. The low- and high-noise profiles varied additive state noise. The jitter profile applied a small symmetric random perturbation to the slow matrix at stage boundaries, testing sensitivity to mild structural fluctuations in the learned operator. The synapse-loss profile additionally introduced random edge deletion, testing sensitivity to mild loss of learned support. The two classical controls were calibrated on a held-out subset of seeds using a non-recall penalty that rewarded selective past- and new-stage responses while penalizing rest activity and saturation. All downstream contrasts were evaluated only after those settings and the chosen support level were frozen. The quantum-like models were not separately tuned in this way. 
This calibration choice was intended to avoid trivial under-tuning of the classical controls, but the comparison should still be read as a matched control analysis under fixed benchmark settings rather than as an exhaustive search over all possible classical baselines. 

The perturbation profiles test whether the same memory mechanism remains interpretable under noise, structural jitter, or mild synaptic disruption. The exact executed perturbation-profile settings are listed in Table~\ref{tab:profile_params}, with the calibration split and workflow order described in Supplementary Section~\ref{sec:supp_pipeline}. The primary recall metric was the trapezoidal AUC over the \(A\)-channel during the final A-cued recall stage: 
\begin{equation}
\mathrm{AUC}_{\mathrm{recallA}}=\sum_{k=k_{\mathrm{start}}}^{k_{\mathrm{end}}-1}\frac{q_A(k)+q_A(k+1)}{2}\,\Delta k,
\end{equation}
where \(q_X(k)=|\langle S_X,\psi(k)\rangle|^2\) for \(X\in\{A,B,C\}\) in the quantum-like and real-valued no-phase models, and \(q_X(k)=p(k)\cdot S_X\) for the Markov-rate control. This quantity measures recovery of the previously learned \(A\) pattern during the final recall stage. Because the integration step and logging interval are fixed across all conditions, AUC values are reported in logged-step units and are used only for matched within-benchmark comparisons.
Auxiliary threshold settings, including \(q^\star\) and the past-anchor threshold, are reported in Table~\ref{tab:sweep_params}.

The stage-structure score operationalizes the staged task logic introduced above. It is a benchmark-specific summary rather than a direct biological observable. Let \(\mathrm{stage}_X\) denote the mean value of channel \(q_X\) over the named stage; for example, \(\mathrm{past}_A\) is the mean \(A\)-channel value during the past stage and \(\mathrm{new}_C\) is the mean \(C\)-channel value during the new stage.
\begin{equation}
\begin{aligned}
S_{\mathrm{stage}} =\;& (\mathrm{past}_A + \mathrm{past}_B - 2\,\mathrm{past}_C) \\
&+ (2\,\mathrm{new}_C - \mathrm{new}_A - \mathrm{new}_B) \\
&+ (2\,\mathrm{recallA}_A - \mathrm{recallA}_B - \mathrm{recallA}_C) \\
&- 1.5\!\left(\max\{\mathrm{rest1}_A,\mathrm{rest1}_B,\mathrm{rest1}_C\}
+\max\{\mathrm{rest2}_A,\mathrm{rest2}_B,\mathrm{rest2}_C\}\right),
\end{aligned}
\end{equation}
where the first line rewards selective expression of the intended \(A/B\) channels during the past stage, the second line rewards preferential expression of the novel $C$ channel during the new stage, the third line rewards selective final $A$-cued recall, and the final penalty suppresses spurious rest activity. Positive values indicate stronger alignment with the intended stage structure.

To quantify context-sensitive order dependence, we froze the post-rest2 state and effective matrix and compared short $A\!\rightarrow\!C$ and $C\!\rightarrow\!A$ probe sequences. Order asymmetry was then summarized as
\begin{equation}
\Delta_{\mathrm{order}} = \big|A_{\mathrm{after}\,AC}-A_{\mathrm{after}\,CA}\big| + \big|C_{\mathrm{after}\,AC}-C_{\mathrm{after}\,CA}\big|,
\end{equation}
 where \(A_{\mathrm{after}\,AC}\) and \(C_{\mathrm{after}\,AC}\) denote the final \(A\)- and \(C\)-channel readouts after the \(A\!\rightarrow\!C\) probe sequence, and \(A_{\mathrm{after}\,CA}\) and \(C_{\mathrm{after}\,CA}\) denote the corresponding final readouts after the reversed \(C\!\rightarrow\!A\) sequence. 
Larger values indicate stronger dependence on probe order at a fixed post-learning state. Together, recall AUC, stage structure, and order asymmetry define the multi-objective benchmark. 

\subsection{Factorial contrasts and paired effects}

All formal comparisons were performed on evaluation seeds paired within matched conditions. Model comparisons contrasted the full quantum-like model with the no-plasticity quantum-like ablation, the real-valued no-phase control, and the Markov-rate control. The full quantum-like model and both classical controls spanned the full \(2^3\) plasticity grid in both weak-support modes, whereas the no-plasticity quantum-like ablation served as the all-off ablation reference. Backbone comparisons contrasted support on versus off within matched profile, model, and plasticity conditions. 

A factorial main effect was defined as the average change in a metric when one binary plasticity mechanism was switched on rather than off, averaging over all settings of the other two mechanisms. 
Pairwise interactions measured departures from additivity between two plasticity mechanisms. Because all contrasts were paired within matched evaluation conditions, the reported mean differences and paired \(d_z\) values should be interpreted as within-condition effects rather than as independent-group summaries. Descriptive profile-level overview plots are used only for orientation; the formal claims in the text come from these paired comparisons. 
Explicit formulas for factorial effects, pairwise interactions, and paired effect sizes are provided in Supplementary Section~\ref{sec:supp_paired_contrasts}.

\section{Results}\label{sec:results}

The results follow the benchmark sequence: weak-support selection, profile-resolved dynamic checks, factorial plasticity decomposition, and matched model-family comparison. Formal interpretation is based on paired matched-condition contrasts computed after support selection and control calibration were frozen. Profile-level views are used to check whether the paired contrasts reflect coherent trajectory-level behavior rather than isolated scalar effects. Supporting sweep traces and diagnostic panels are provided in Supplementary Figures~\ref{fig:supp_overlay_dashboard}--\ref{fig:supp_model_family_order_overview}, additional quantitative notes in Supplementary Section~\ref{sec:supp_results_notes}, and expanded result tables in Supplementary Section~\ref{sec:supp_tables}.

\subsection{Selection of the weak-support operating point}
The support-selection analysis identified a narrow, interior, and non-rescuing operating regime. The structural floor was swept over the zero-floor reference and nonzero values from \(10^{-15}\) to \(10^{-6}\) under a sparse modular scaffold. The operating point was selected conservatively: the full quantum-like model had to improve without inducing a comparable gain in the no-plasticity quantum-like ablation on the screening metrics. Under this criterion, the selected support level was the interior point \(\epsilon=10^{-10}\), not the largest floor value tested. At the aggregate level, the full-model recall AUC increased modestly from \(768.48\) at \(\epsilon=0\) to \(777.10\) at \(\epsilon=10^{-10}\), while the no-plasticity ablation remained effectively unchanged. The sweep therefore supports two bounded conclusions: useful support lies in a narrow non-monotonic regime, and the selected floor does not behave like a hidden rescue mechanism for the ablation reference.

\begin{figure}[t]
\centering
\includegraphics[width=\linewidth]{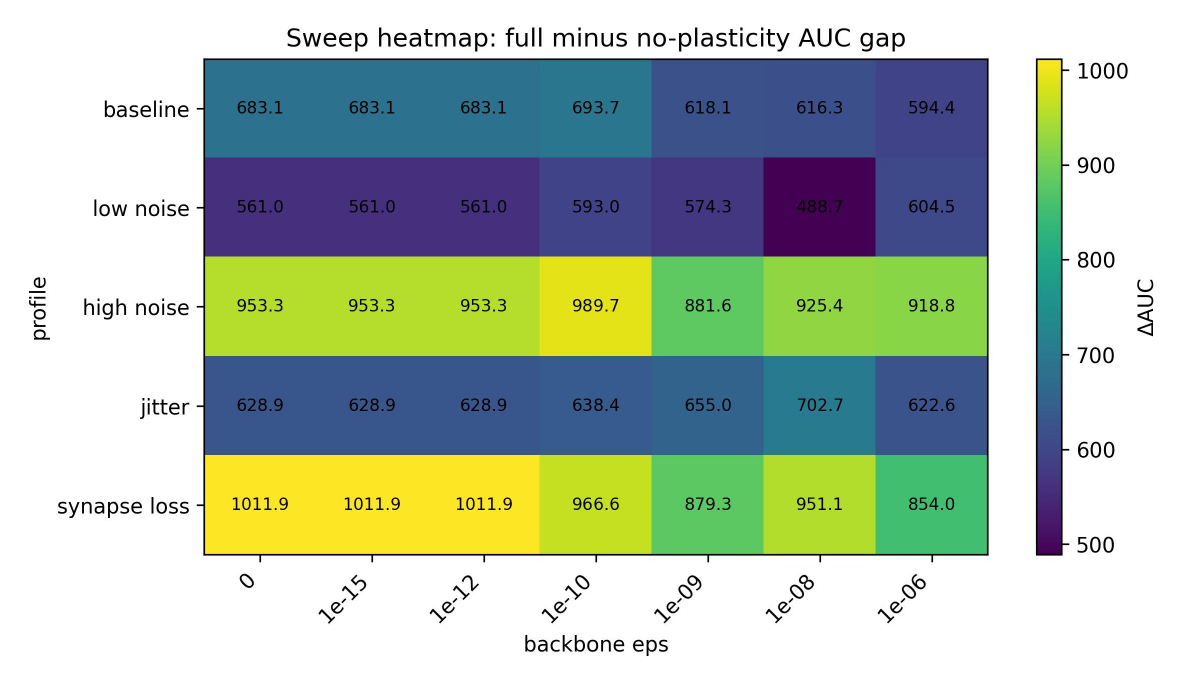}
\caption{Profile-resolved weak-support sweep viewed through the recall gap between the full quantum-like model and its no-plasticity ablation. Each cell reports the mean $\mathrm{AUC}_{\mathrm{recallA}}$ difference at one profile-\(\epsilon\) combination. The useful regime is again interior rather than monotonic: the largest gaps concentrate around \(\epsilon=10^{-10}\) for the high-noise and synapse-loss profiles, while low-noise and jitter exhibit narrower gains. This heatmap therefore complements the scalar selection score by showing where the selected weak-support floor remains beneficial without materially rescuing the ablation.}
\label{fig:backbone_sweep_gap}
\end{figure}

In the frozen factorial run, enabling the weak structural floor had only small aggregate effects. For the full quantum-like model, the mean shift was \(\Delta\mathrm{AUC}_{\mathrm{recallA}}=-1.22\), with corresponding shifts of \(\Delta_{\mathrm{order}}=-9.5\times10^{-5}\) and \(\Delta S_{\mathrm{stage}}=+8.4\times10^{-4}\). Figure~\ref{fig:backbone_sweep_gap} shows that the support effect was not distributed uniformly across profiles: the largest full-minus-ablation recall gaps occurred in the high-noise and synapse-loss conditions, whereas low-noise and jitter showed narrower gains. In the present benchmark, the scaffold is therefore best interpreted as a weak stabilizer of the operating point rather than as the primary driver of recall.

\subsection{Profile-resolved mechanistic traces at the selected operating point}

Profile-resolved trajectories were used to check that the selected operating point produced interpretable dynamics rather than only a favorable aggregate score. At \(\epsilon=10^{-10}\), the full quantum-like model produced substantial final recall across the baseline, low-noise, high-noise, jitter, and synapse-loss profiles, with mean recall AUCs of \(694.6\), \(593.8\), \(990.5\), \(639.2\), and \(967.4\), respectively. By contrast, the no-plasticity quantum-like ablation remained near floor, with recall AUC \(0.827\) in each profile. Thus, the selected operating point does not merely improve an aggregate score; it preserves a large full-minus-ablation recall separation across the perturbation set.

The trajectory-level diagnostics clarify how this separation is stage-specific. In the full model, \(q_A(k)\) remains stage selective: it is activated during the past stage, suppressed through the intervening stages, and rebuilt during the final A-cued recall stage. The high-noise and synapse-loss profiles show the largest late recall responses, consistent with Fig.~\ref{fig:backbone_sweep_gap}. In the no-plasticity ablation, the same \(q_A(k)\) trajectory remains comparatively flat, indicating that the weak structural floor alone does not recover the cued memory. Expanded trajectory, compactness, stage-resolved, and profile-level summaries are provided in Supplementary Figures~\ref{fig:supp_overlay_dashboard}--\ref{fig:supp_profile_metric_summary}.

\subsection{Plasticity effects under the selected weak-support regime}
After the weak-support operating point was fixed, the factorial plasticity analysis identified homeostatic stabilization as the clearest first-order contributor to recall. The averages reported here are computed over the 10 evaluation condition cells defined by the five perturbation profiles and the two support modes. Across these conditions, homeostatic plasticity produced the largest positive main effect on recall AUC in the full quantum-like model (mean main effect \(+157.6\); mean paired \(d_z=1.10\)), structural plasticity was positive but weaker (\(+36.9\); \(d_z=0.17\)), and heterosynaptic plasticity was slightly negative on average at first order (\(-20.6\); \(d_z=-0.08\)). At the level of first-order recall effects, regulated stabilization is therefore the most reliable contributor, whereas rewiring and competition matter more conditionally.

Pairwise interactions refined this picture. The strongest positive interaction involved heterosynaptic and structural plasticity, whereas the homeostatic--heterosynaptic interaction was weaker and slightly negative on average. Thus, the slightly negative first-order heterosynaptic effect does not mean that heterosynaptic plasticity is uniformly detrimental; rather, it becomes useful in the paired context created by structural reconfiguration. Homeostatic stabilization therefore provides the clearest first-order recall benefit, while structural and heterosynaptic mechanisms contribute more conditionally.

\begin{figure*}[t]
\centering
\includegraphics[width=\linewidth]{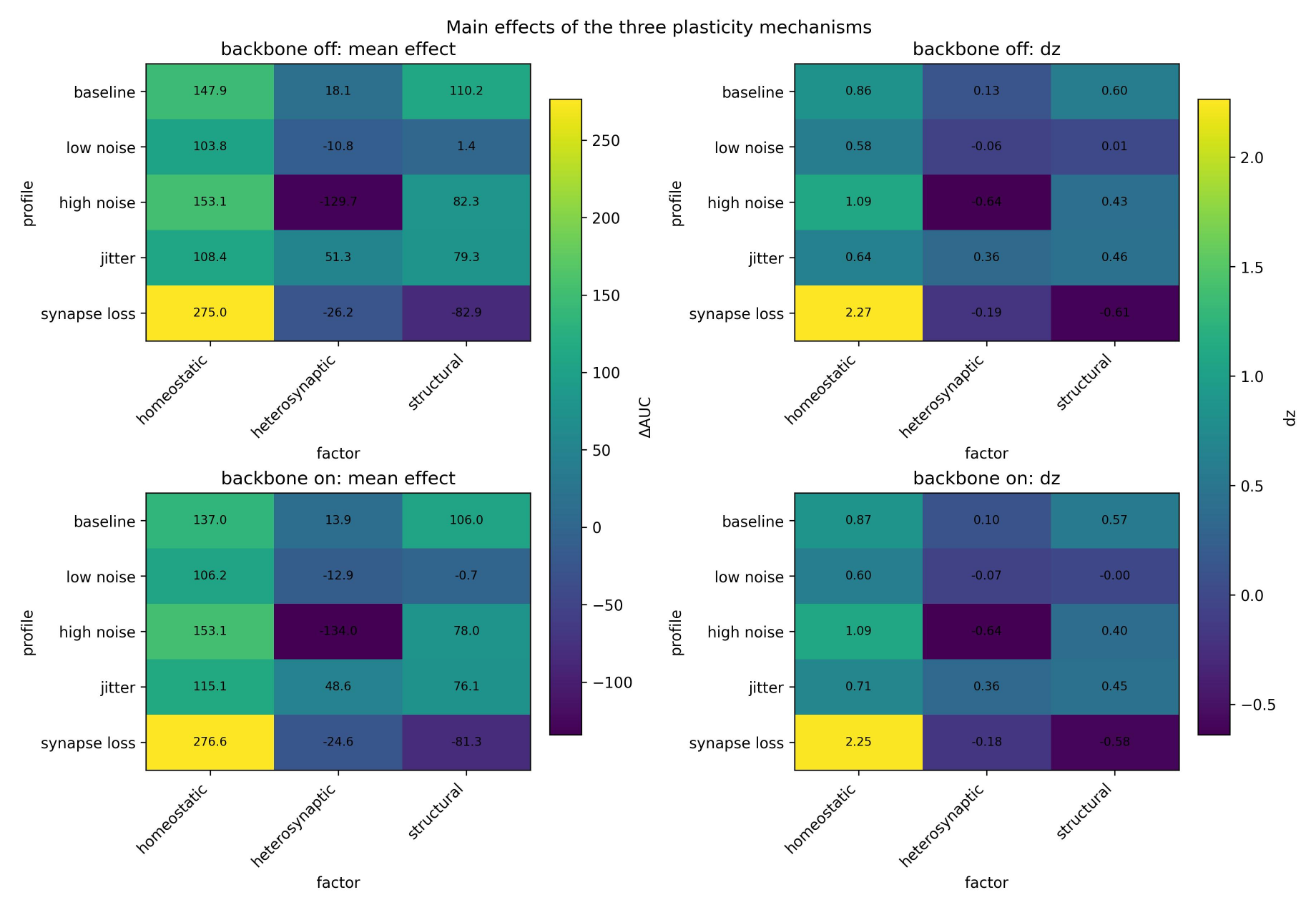}
\caption{Main effects of homeostatic, heterosynaptic, and structural plasticity on recall performance. For each factor, the main effect was computed by averaging over all factorial conditions in which that factor was on and subtracting the average over conditions in which it was off. The corresponding paired $d_z$ effect sizes are shown alongside the raw mean contrasts. Top and bottom rows show support-off and support-on regimes, respectively.}
\label{fig:main_effects_auc}
\end{figure*}

Figure~\ref{fig:main_effects_auc} summarizes the first-order effects. This ranking should be read as benchmark-specific for the recall objective reported here, not as a universal ordering of biological importance.

\subsection{Matched comparison with classical controls}
The matched model comparison clarifies what the quantum-like formulation contributes in this benchmark. If the benchmark were evaluated only by raw recall AUC, the Markov-rate control would be the strongest classical comparator. In matched conditions, however, the Markov-rate control is stronger mainly on raw recall AUC, whereas the quantum-like model is stronger on order asymmetry and stage structure. Relative to the real-valued no-phase control, the quantum-like model is favorable on average across all three benchmark summaries.

\begin{figure}[t]
    \centering
    \includegraphics[width=0.96\linewidth]{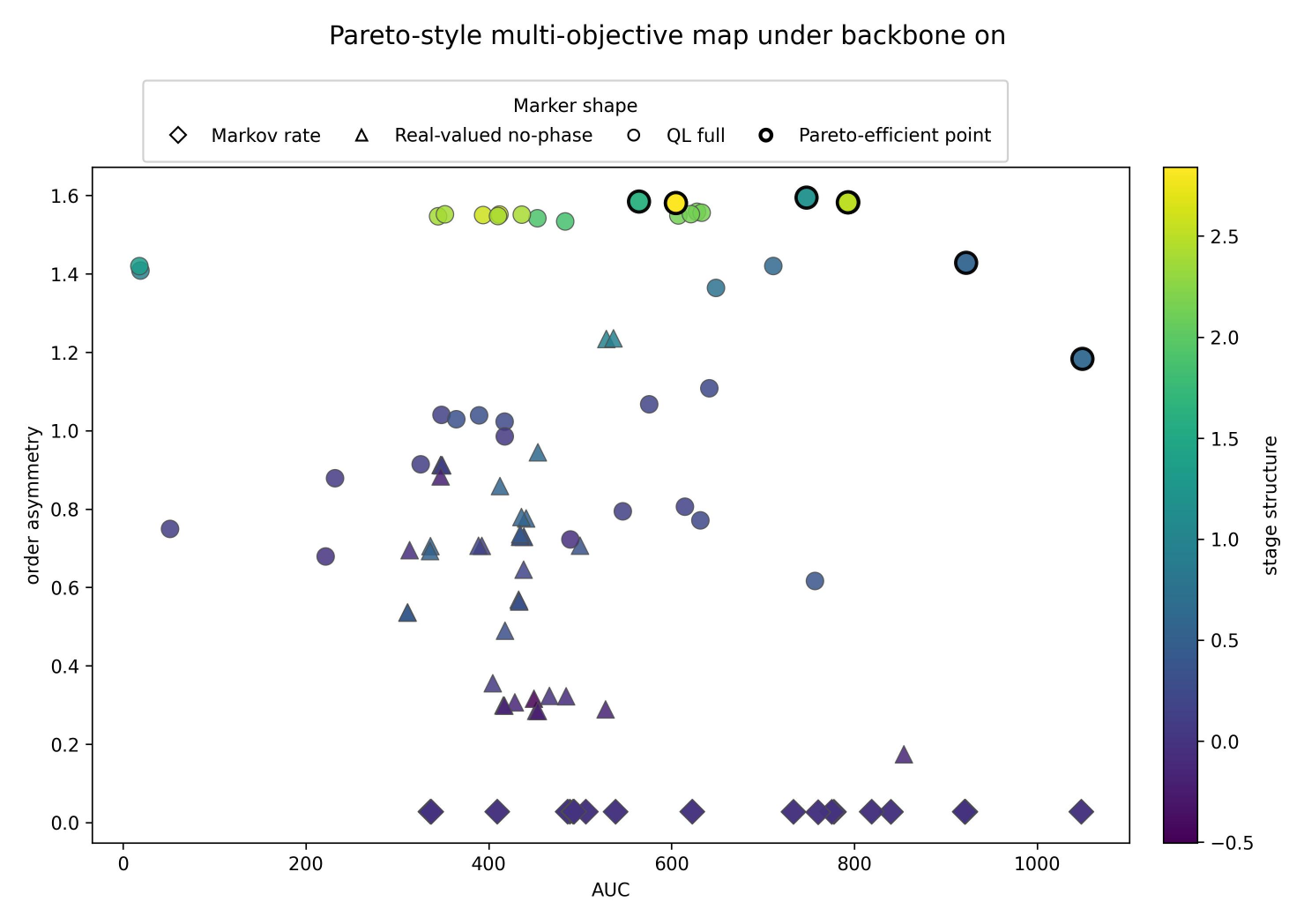}
    \caption{Backbone-on multi-objective comparison across the full evaluation cloud. Each point represents one evaluated model--profile--plasticity cell, plotted as recall AUC versus order asymmetry, with stage-structure score encoded by color. Marker shape denotes model family. Heavy outlines indicate Pareto-efficient cells under the three-objective benchmark. Within the evaluated backbone-on grid and using point-estimate summaries, the Pareto-efficient cells belong to the quantum-like model with plasticity.}
    \label{fig:pareto_multiobjective_main}
\end{figure}

The comparison is therefore not a claim that the quantum-like model maximizes every scalar summary. Rather, it shows that the model families occupy different regions of the joint recall--order--stage-structure benchmark space. Figure~\ref{fig:pareto_multiobjective_main} shows the full backbone-on set of evaluated model--profile--plasticity cells, with recall AUC and order asymmetry on the axes and stage structure encoded by color. A cell is Pareto-efficient if no other evaluated cell is at least as good on all three benchmark objectives and strictly better on at least one of them. Within the evaluated backbone-on grid and using point-estimate summaries, the Pareto-efficient cells belong to the quantum-like model with plasticity.

The high-noise profile is the clearest stress-test regime: in that setting, the quantum-like formulation retains its order-sensitive and stage-structure advantages while also becoming recall-competitive against the Markov-rate control. The strong high-noise recall response should be interpreted as a benchmark-specific perturbation effect, not as a general claim that noise improves memory. In this task, the noise profile appears to interact with the adaptive operator and cueing schedule in a way that can increase late-stage \(A\)-channel recovery, and this effect should be tested in future sensitivity sweeps. Broader control-family factorial summaries support the same interpretation: Markov-rate varies mainly along a recall axis, the real-valued no-phase family reaches an intermediate order-sensitive regime, and the quantum-like family retains the strongest order-sensitive organization across the shared benchmark grid.

\begin{figure}[t]
\centering
\includegraphics[width=\linewidth]{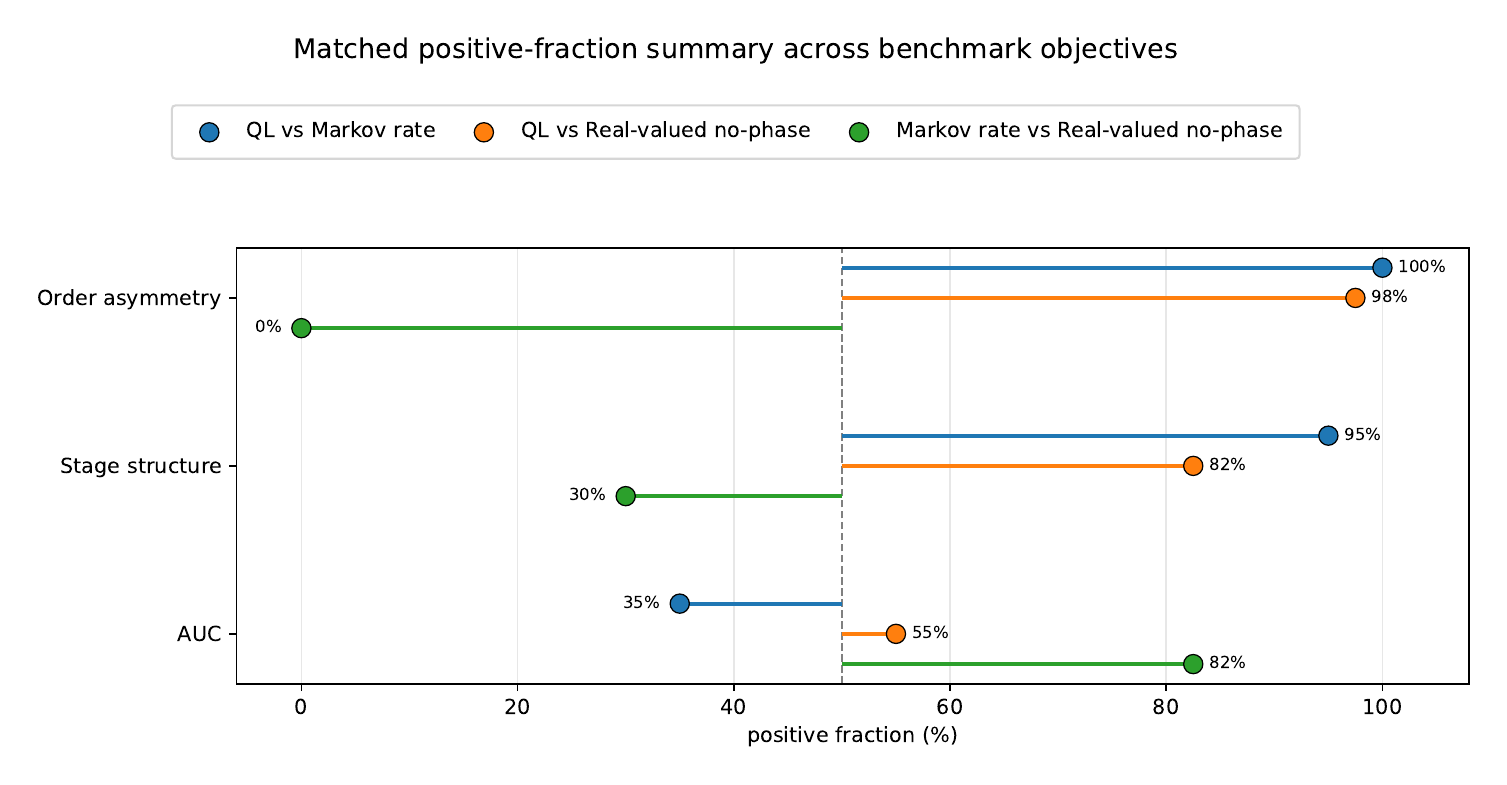}
\caption{Matched positive-fraction comparison across benchmark objectives. For each paired evaluation cell, the figure records whether the first model in each comparison exceeds the second on recall AUC, order asymmetry, or stage-structure score. The vertical dashed line denotes parity at \(50\%\). The quantum-like model is less consistently favored over the Markov-rate control on recall AUC, but is favored on order asymmetry and stage structure; relative to the real-valued no-phase control, it is favored across all three summaries. The Markov-rate control is favored over the real-valued no-phase control primarily on recall AUC, indicating that high scalar recall and context-sensitive temporal structure separate across model classes.}
\label{fig:matched_positive_fraction}
\end{figure}

The inferential takeaway from Fig.~\ref{fig:matched_positive_fraction} is deliberately asymmetric. Against the Markov-rate control, the quantum-like model wins far more consistently on order asymmetry and stage structure than on raw recall AUC, while against the real-valued no-phase control it remains favorable on all three benchmark summaries in aggregate. The direct comparison between the two classical controls further separates scalar recall from temporal-structure sensitivity: the Markov-rate control exceeds the real-valued no-phase control in \(82.5\%\) of matched cells on recall AUC, but only in \(0\%\) on order asymmetry and \(30\%\) on stage structure. Seed-level bootstrap confidence intervals for the principal matched contrasts are reported in Supplementary Table~\ref{tab:principal_seed_ci} and support the same qualitative pattern.

Together, the matched comparisons and the control-family factorial checks show that the classical controls do not reproduce the quantum-like model's joint benchmark profile. The appropriate justification for the quantum-like model is therefore not universal scalar superiority, but its more consistent combination of substantial recall, order sensitivity, and positive stage structure under matched perturbation-resolved comparison. Table~\ref{tab:benchmark_synthesis} summarizes how the three main result blocks answer the study's central inferential questions.

\begingroup
\footnotesize
\renewcommand{\arraystretch}{1.25}
\setlength{\tabcolsep}{3pt}
\setlength{\LTleft}{0pt}
\setlength{\LTright}{0pt}

\begin{longtable}{
>{\raggedright\arraybackslash}p{0.18\textwidth}
>{\raggedright\arraybackslash}p{0.16\textwidth}
>{\raggedright\arraybackslash}p{0.38\textwidth}
>{\raggedright\arraybackslash}p{0.20\textwidth}
}
    \caption{Benchmark synthesis across the three inferential questions of the study. The table condenses the main result logic from the primary figures and supporting quantitative tables.}
    \label{tab:benchmark_synthesis}\\

    \toprule
    Inferential question & Main evidence & Quantitative summary & Interpretation \\
    \midrule
    \endfirsthead

    \toprule
Inferential question & Main evidence & Quantitative summary & Interpretation \\
\midrule
\endhead
    
    Which support regime remains useful once scaffold rescue is excluded? & Fig.~\ref{fig:backbone_sweep_gap} & Full-model mean recall AUC rises from $768.48$ at $\epsilon=0$ to $777.10$ at $\epsilon=10^{-10}$, while the no-plasticity ablation remains fixed at $0.827267$. In the frozen factorial benchmark, backbone-on minus backbone-off shifts remain small on average ($\Delta\mathrm{AUC}=-1.22$, \(\Delta_{\mathrm{order}}=-9.5\times10^{-5}\), $\Delta S_{\mathrm{stage}}=+8.4\times10^{-4}$). & Weak structure acts as a conservative stabilizer of an operating point rather than as a hidden rescue mechanism. \\
    \addlinespace
    Which plasticity mechanisms carry the recall effect once support is fixed? & Fig.~\ref{fig:main_effects_auc} & Homeostatic plasticity shows the largest positive first-order recall effect ($+157.6$, $d_z=1.10$), structural plasticity is weaker ($+36.9$, $d_z=0.17$), and heterosynaptic plasticity is slightly negative on average at first order ($-20.6$, $d_z=-0.08$) but strongly positive in interaction with structural plasticity ($+258.9$, $d_z=0.75$). & Recall is carried primarily by regulated plasticity, with stabilization dominant at first order and competitive or rewiring mechanisms contributing conditionally. \\
    \addlinespace
    Why use the quantum-like formulation once classical controls are matched on the same task? & Figs.~\ref{fig:pareto_multiobjective_main} and \ref{fig:matched_positive_fraction} & Versus Markov-rate, the quantum-like model with plasticity has lower mean recall AUC ($\Delta\mathrm{AUC}=-154.0$) but higher mean order asymmetry and stage structure (\(\Delta_{\mathrm{order}}=+1.236\), $\Delta S_{\mathrm{stage}}=+1.208$); the corresponding positive fractions are $35\%$, $100\%$, and $95\%$. Versus real no-phase, it is ahead on all three metrics on average ($+79.1$, $+0.629$, $+0.982$). & Markov-rate is strongest on scalar recall, whereas the quantum-like model more consistently preserves the benchmark's context-sensitive multi-objective balance. \\
    
    \bottomrule
\end{longtable}
\endgroup

\section{Discussion}\label{sec:discussion}

The main claim of this study is intentionally modest. We do not argue that the model demonstrates a general quantum advantage, or that it establishes microscopic quantum computation in neural tissue. Instead, the work should be read as a benchmark for a specific kind of memory problem: how a system can recall a previous context after intervening experience, while still preserving the order and structure of the task.

Three conclusions emerge from the conservative design. First, weak structural support can help define a stable operating regime, but it does not by itself solve the task. Second, once that regime is fixed, the contributions of different plasticity mechanisms can be separated. Third, the quantum-like formulation is useful here because it preserves recall together with sequential context sensitivity, not because it maximizes every single performance metric.

The support-selection result is important because it addresses a common concern in structured memory models. If the fixed scaffold were too strong, it could quietly carry the task and make the learned dynamics look more successful than they really are. That was not the case here. The selected support level was a weak interior point rather than the largest value tested, and the no-plasticity ablation was not rescued by adding this support. The scaffold therefore behaves more like a stabilizing constraint than like a hidden solution to the recall problem.

The ablation results reinforce this interpretation. The no-plasticity model retained some stage separation and order sensitivity, but it did not produce meaningful final recall. In other words, preserving the shape of the task is not enough; the system must also recover the relevant prior context. This is why the benchmark is most informative when recall, stage structure, and order sensitivity are considered together. Within that joint view, homeostatic plasticity provided the clearest first-order recall benefit, consistent with the broader role of stabilizing feedback in plastic neural systems \cite{turrigiano2012}. Structural plasticity gave a smaller average benefit, and heterosynaptic plasticity was most useful when combined with structural reconfiguration.

The comparison with classical controls further clarifies the role of the quantum-like formulation. A scalar recall score alone would give an incomplete picture, because the Markov-rate control is highly competitive on raw recall but shows negligible order asymmetry and weak or negative stage-structure scores. The quantum-like model does not always dominate on recall alone, but it better preserves recall together with the staged, path-dependent organization that the task was designed to test. Its value in this benchmark is therefore not scalar superiority, but preservation of a more complete memory profile.

This framing also keeps the claims limited. The results do not rule out all possible classical explanations, nor do they show that quantum-like dynamics are universally better for memory modeling. A broader classical model family, a different calibration strategy, or additional task families could reduce or remove the observed advantage. The present contribution is narrower: within the tested benchmark and matched controls, the quantum-like model preserves a joint recall--context--structure trade-off that the classical controls do not reproduce as consistently.

Several limitations follow from this interpretation. The framework is computational and abstract, so the plasticity mechanisms should be understood as operational rules rather than direct biological mechanisms. The order-asymmetry score is a proxy for context sensitivity, not evidence of noncommutativity in neural tissue. The benchmark also uses one staged task family and a limited set of controls. Future work should therefore test additional task structures, broader classical baselines, and formal Pareto-style comparisons across recall, order sensitivity, and stage structure \cite{marler2004survey,deb2002nsga2}.

A natural biological extension would be to replace the present operational plasticity rules with experimentally constrained learning rules. Behavioral timescale synaptic plasticity (BTSP) is one candidate because it links rapid synaptic change, dendritic plateau events, and hippocampal representation formation on behaviorally relevant timescales \cite{magee2026btsp,madar2025btsp}. In this benchmark, BTSP-inspired rules would not change the central claim, but would test whether the same balance between weak structure, regulated plasticity, and context-sensitive recall persists under more biologically grounded update dynamics.

Overall, the limitations narrow the scope of the message without weakening it. Robust staged recall in this framework depends on a balance between weak support, regulated plasticity, and multi-objective evaluation. Homeostatic stabilization provides the clearest first-order benefit, while the quantum-like formulation earns its place by preserving the sequential, context-sensitive structure that the task was designed to probe.

\section{Conclusion}\label{sec:conclusion}

This study introduced a reproducible staged benchmark for associative recall under weak structural support, adaptive plasticity, perturbation, and matched classical comparison. Three conclusions follow. First, useful structural support remained weak. The selected support level, $\epsilon = 10^{-10}$, acted as a stabilizer of the operating regime rather than as a hidden rescue mechanism for the no-plasticity ablation. Second, successful recall was driven mainly by adaptive plasticity rather than by the scaffold itself. Homeostatic stabilization provided the clearest first-order benefit, structural plasticity contributed more modestly, and heterosynaptic plasticity was most useful through interactions with structural reconfiguration. Third, the quantum-like formulation was supported not by universal superiority on a single recall score, but by its more consistent preservation of recall together with order sensitivity and stage structure. Overall, the results identify weak support, regulated plasticity, and multi-objective evaluation as the key ingredients of this benchmark, and provide a basis for extending this approach to additional task families and disease-relevant perturbation regimes.

\backmatter

\bmhead{Supplementary information}


The Supplementary Information is included below in this arXiv version.

\bmhead{Acknowledgements}

During manuscript preparation, the authors used a GPT-assisted language tool to improve clarity and readability. The tool was not used to generate data, run simulations, create results, or make scientific decisions. All GPT-assisted text was critically reviewed, edited, and verified by the authors, who take full responsibility for the content of the manuscript. The authors thank those who visited our poster stand at the Phys10 Third Poster Fair at the University of Waterloo for their valuable questions and discussions.

\section*{Declarations}

\subsection*{Funding}
This research was enabled in part by support provided by the Digital Research Alliance of Canada to T.J.A.C. This research was undertaken in part thanks to funding to T.J.A.C. from the Canada Research Chairs Program (CRC-2022-00204) and the University of Waterloo. L.G. acknowledges support from the University of Waterloo Provost's Program for Interdisciplinary Postdoctoral Scholars.

\subsection*{Competing interests}
The authors have no competing interests to declare that are relevant to the content of this article.

\subsection*{Ethics approval}
Not applicable. This study is computational and did not involve human participants, human data, animals, or animal tissue.

\subsection*{Consent to participate}
Not applicable.

\subsection*{Consent for publication}
Not applicable.

\subsection*{Data and code availability}
The data and code supporting the findings of this study consist of simulation outputs, processed summary tables, figure-generation inputs, and custom simulation, analysis, and plotting scripts. These materials are available from the corresponding author upon reasonable request.

\subsection*{Author contributions}
Y.H.G. and L.G. contributed to the conception and design of the study. Y.H.G. performed the simulations, analysis, and figure generation. L.G. and T.J.A.C. contributed to supervision, interpretation, and scientific framing. T.J.A.C. provided resources and contributed to project administration and funding acquisition. The first draft of the manuscript was prepared by Y.H.G. and L.G. All authors reviewed, edited, and approved the final manuscript.

\bibliography{1.references}

@book{hebb1949organization,
  author    = {Hebb, D. O.},
  title     = {The Organization of Behavior: A Neuropsychological Theory},
  publisher = {Psychology Press},
  address   = {New York},
  year      = {2002},
  note      = {2002 reissue of the original 1949 work.}
}

@book{dayanabbott2001,
  author    = {Dayan, Peter and Abbott, L. F.},
  title     = {Theoretical Neuroscience: Computational and Mathematical Modeling of Neural Systems},
  publisher = {MIT Press},
  address   = {Cambridge, MA},
  year      = {2001}
}

@article{blisslomo1973,
  author  = {Bliss, T. V. P. and L{\o}mo, T.},
  title   = {Long-lasting potentiation of synaptic transmission in the dentate area of the anaesthetized rabbit following stimulation of the perforant path},
  journal = {The Journal of Physiology},
  year    = {1973},
  volume  = {232},
  number  = {2},
  pages   = {331--356},
  doi     = {10.1113/jphysiol.1973.sp010273}
}

@article{hopfield1982,
  author  = {Hopfield, J. J.},
  title   = {Neural networks and physical systems with emergent collective computational abilities},
  journal = {Proceedings of the National Academy of Sciences of the United States of America},
  year    = {1982},
  volume  = {79},
  number  = {8},
  pages   = {2554--2558},
  doi     = {10.1073/pnas.79.8.2554}
}

@article{oja1982,
  author  = {Oja, Erkki},
  title   = {A simplified neuron model as a principal component analyzer},
  journal = {Journal of Mathematical Biology},
  year    = {1982},
  volume  = {15},
  number  = {3},
  pages   = {267--273},
  doi     = {10.1007/BF00275687}
}

@article{bi1998,
  author  = {Bi, Guo-Qiang and Poo, Mu-Ming},
  title   = {Synaptic modifications in cultured hippocampal neurons: dependence on spike timing, synaptic strength, and postsynaptic cell type},
  journal = {The Journal of Neuroscience},
  year    = {1998},
  volume  = {18},
  number  = {24},
  pages   = {10464--10472},
  doi     = {10.1523/JNEUROSCI.18-24-10464.1998}
}

@article{turrigiano2012,
  author  = {Turrigiano, Gina},
  title   = {Homeostatic synaptic plasticity: local and global mechanisms for stabilizing neuronal function},
  journal = {Cold Spring Harbor Perspectives in Biology},
  year    = {2012},
  volume  = {4},
  number  = {1},
  pages   = {a005736},
  doi     = {10.1101/cshperspect.a005736}
}

@article{chistiakova2015,
  author  = {Chistiakova, Marina and Bannon, Nicholas M. and Chen, Jen-Yung and Bazhenov, Maxim and Volgushev, Maxim},
  title   = {Homeostatic role of heterosynaptic plasticity: models and experiments},
  journal = {Frontiers in Computational Neuroscience},
  year    = {2015},
  volume  = {9},
  pages   = {89},
  doi     = {10.3389/fncom.2015.00089}
}

@article{jenks2021,
  author  = {Jenks, Kyle R. and Tsimring, Katya and Ip, Jacque Pak Kan and Zepeda, Jose C. and Sur, Mriganka},
  title   = {Heterosynaptic plasticity and the experience-dependent refinement of developing neuronal circuits},
  journal = {Frontiers in Neural Circuits},
  year    = {2021},
  volume  = {15},
  pages   = {803401},
  doi     = {10.3389/fncir.2021.803401}
}

@article{holtmaat2009,
  author  = {Holtmaat, Anthony and Svoboda, Karel},
  title   = {Experience-dependent structural synaptic plasticity in the mammalian brain},
  journal = {Nature Reviews Neuroscience},
  year    = {2009},
  volume  = {10},
  number  = {9},
  pages   = {647--658},
  doi     = {10.1038/nrn2699}
}

@article{breakspear2017,
  author  = {Breakspear, Michael},
  title   = {Dynamic models of large-scale brain activity},
  journal = {Nature Neuroscience},
  year    = {2017},
  volume  = {20},
  number  = {3},
  pages   = {340--352},
  doi     = {10.1038/nn.4497}
}

@book{busemeyerbruza2012,
  author    = {Busemeyer, Jerome R. and Bruza, Peter D.},
  title     = {Quantum Models of Cognition and Decision},
  publisher = {Cambridge University Press},
  address   = {Cambridge},
  year      = {2012}
}

@article{pothos2013,
  author  = {Pothos, Emmanuel M. and Busemeyer, Jerome R.},
  title   = {Can quantum probability provide a new direction for cognitive modeling?},
  journal = {Behavioral and Brain Sciences},
  year    = {2013},
  volume  = {36},
  number  = {3},
  pages   = {255--274},
  doi     = {10.1017/S0140525X12001525}
}

@article{wang2014,
  author  = {Wang, Zheng and Solloway, Tyler and Shiffrin, Richard M. and Busemeyer, Jerome R.},
  title   = {Context effects produced by question orders reveal quantum nature of human judgments},
  journal = {Proceedings of the National Academy of Sciences of the United States of America},
  year    = {2014},
  volume  = {111},
  number  = {26},
  pages   = {9431--9436},
  doi     = {10.1073/pnas.1407756111}
}

@article{newman2006,
  author  = {Newman, M. E. J.},
  title   = {Modularity and community structure in networks},
  journal = {Proceedings of the National Academy of Sciences of the United States of America},
  year    = {2006},
  volume  = {103},
  number  = {23},
  pages   = {8577--8582},
  doi     = {10.1073/pnas.0601602103}
}

@inproceedings{ng2001spectral,
  author    = {Ng, Andrew Y. and Jordan, Michael I. and Weiss, Yair},
  title     = {On Spectral Clustering: Analysis and an algorithm},
  booktitle = {Advances in Neural Information Processing Systems 14},
  year      = {2001},
  pages     = {849--856}
}

@article{scholes2024qlstates,
  author  = {Scholes, Gregory D.},
  title   = {Quantum-like states on complex synchronized networks},
  journal = {Proceedings of the Royal Society A: Mathematical, Physical and Engineering Sciences},
  year    = {2024},
  volume  = {480},
  number  = {2295},
  pages   = {20240209},
  doi     = {10.1098/rspa.2024.0209}
}

@article{scholes2026dynamics,
  author  = {Scholes, Gregory D.},
  title   = {Dynamics in an emergent quantum-like state space generated by a nonlinear classical network},
  journal = {Physica Scripta},
  year    = {2026},
  volume  = {101},
  number  = {5},
  pages   = {055205},
  doi     = {10.1088/1402-4896/ae3d90}
}

@article{amati2025qlbits,
  author  = {Amati, Graziano and Scholes, Gregory D.},
  title   = {Quantum information with quantumlike bits},
  journal = {Physical Review A},
  year    = {2025},
  volume  = {111},
  number  = {6},
  pages   = {062203},
  doi     = {10.1103/PhysRevA.111.062203}
}

@article{amati2025encoding,
  title = {Encoding quantumlike information in classical synchronizing dynamics},
  author = {Amati, Graziano and Scholes, Gregory D.},
  journal = {Phys. Rev. A},
  volume = {112},
  issue = {3},
  pages = {032423},
  numpages = {12},
  year = {2025},
  month = {Sep},
  publisher = {American Physical Society},
  doi = {10.1103/nvdp-sr11},
}

@article{marler2004survey,
  author = {Marler, R. T. and Arora, J. S.},
  title = {Survey of multi-objective optimization methods for engineering},
  journal = {Structural and Multidisciplinary Optimization},
  year = {2004},
  volume = {26},
  number = {6},
  pages = {369--395},
  doi = {10.1007/s00158-003-0368-6}
}

@article{deb2002nsga2,
  author = {Deb, Kalyanmoy and Pratap, Amrit and Agarwal, Sameer and Meyarivan, T.},
  title = {A fast and elitist multiobjective genetic algorithm: {NSGA-II}},
  journal = {IEEE Transactions on Evolutionary Computation},
  year = {2002},
  volume = {6},
  number = {2},
  pages = {182--197},
  doi = {10.1109/4235.996017}
}

@article{Geirhos2020Shortcut,
  author  = {Geirhos, Robert and Jacobsen, J{\"o}rn-Henrik and Michaelis, Claudio and Zemel, Richard and Brendel, Wieland and Bethge, Matthias and Wichmann, Felix A.},
  title   = {Shortcut Learning in Deep Neural Networks},
  journal = {Nature Machine Intelligence},
  year    = {2020},
  volume  = {2},
  number  = {11},
  pages   = {665--673},
  doi     = {10.1038/s42256-020-00257-z}
}

@article{Lever2016ModelSelection,
  author  = {Lever, Jake and Krzywinski, Martin and Altman, Naomi},
  title   = {Model selection and overfitting},
  journal = {Nature Methods},
  year    = {2016},
  volume  = {13},
  number  = {9},
  pages   = {703--704},
  doi     = {10.1038/nmeth.3968}
}

@article{Weber2019Benchmarking,
  author  = {Weber, Lukas M. and Saelens, Wouter and Cannoodt, Robrecht and
             Soneson, Charlotte and Hapfelmeier, Alexander and Gardner, Paul P. and
             Boulesteix, Anne-Laure and Saeys, Yvan and Robinson, Mark D.},
  title   = {Essential guidelines for computational method benchmarking},
  journal = {Genome Biology},
  year    = {2019},
  volume  = {20},
  number  = {1},
  pages   = {125},
  doi     = {10.1186/s13059-019-1738-8}
}

@article{Bowles2024SubtleArt,
  author        = {Bowles, Joseph and Ahmed, Shahnawaz and Schuld, Maria},
  title         = {Better than classical? The subtle art of benchmarking quantum machine learning models},
  journal       = {arXiv},
  year          = {2024},
  eprint        = {2403.07059},
  archivePrefix = {arXiv},
  primaryClass  = {quant-ph}
}

@article{zenke2017temporal,
  author = {Zenke, Friedemann and Gerstner, Wulfram and Ganguli, Surya},
  title = {The temporal paradox of Hebbian learning and homeostatic plasticity},
  journal = {Current Opinion in Neurobiology},
  year = {2017},
  volume = {43},
  pages = {166--176},
  doi = {10.1016/j.conb.2017.03.015}
}

@article{toyoizumi2014dynamic,
  author = {Toyoizumi, Taro and Kaneko, Megumi and Stryker, Michael P. and Miller, Kenneth D.},
  title = {Modeling the dynamic interaction of Hebbian and homeostatic plasticity},
  journal = {Neuron},
  year = {2014},
  volume = {84},
  number = {2},
  pages = {497--510},
  doi = {10.1016/j.neuron.2014.09.036}
}

@article{butz2009activity,
  author  = {Butz, Markus and W{\"o}rg{\"o}tter, Florentin and van Ooyen, Arjen},
  title   = {Activity-dependent structural plasticity},
  journal = {Brain Research Reviews},
  year    = {2009},
  volume  = {60},
  number  = {2},
  pages   = {287--305},
  doi     = {10.1016/j.brainresrev.2008.12.023}
}

@article{ausilio2026memory,
  title={Memory preservation and cooperative shielding in complex quantum networks},
  author={Ausilio, Simone and Borgonovi, Fausto and Celardo, Giuseppe Luca and Yago Malo, Jorge and Chiofalo, Maria Luisa},
  journal={Physical Review B},
  volume={113},
  number={6},
  pages={064201},
  year={2026},
  publisher={APS},
  doi     = {10.1103/dt4n-8sqg}
}

@article{magee2026btsp,
  author  = {Magee, Jeffrey C},
  title   = {Behavioral timescale synaptic plasticity: properties, elements and functions},
  journal = {Nature Neuroscience},
  year    = {2026},
  volume  = {29},
  pages={1--15},
  publisher={Nature Publishing Group US New York},
  doi     = {10.1038/s41593-026-02214-2}
}

@article{madar2025btsp,
  author  = {Madar, Antoine D. and Milstein, Aaron D. and O'Dell, Thomas J. and Jain, Anant and Clopath, Claudia and Sheffield, Mark E. J.},
  title   = {Behavioral Timescale Synaptic Plasticity: A Burst in the Field of Learning and Memory},
  journal = {The Journal of Neuroscience},
  year    = {2025},
  volume  = {45},
  number  = {46},
  publisher={Society for Neuroscience},
  doi     = {10.1523/JNEUROSCI.1332-25.2025}
}



\clearpage
\phantomsection
\section*{Supplementary Information}
\addcontentsline{toc}{section}{Supplementary Information}
\label{sec:supplementary}

\setcounter{section}{0}
\setcounter{subsection}{0}
\setcounter{figure}{0}
\setcounter{table}{0}
\setcounter{equation}{0}

\renewcommand{\thesection}{S\arabic{section}}
\renewcommand{\thesubsection}{S\arabic{section}.\arabic{subsection}}
\renewcommand{\thefigure}{S\arabic{figure}}
\renewcommand{\thetable}{S\arabic{table}}
\renewcommand{\theequation}{S\arabic{equation}}

\renewcommand{\theHsection}{S\arabic{section}}
\renewcommand{\theHsubsection}{S\arabic{section}.\arabic{subsection}}
\renewcommand{\theHfigure}{S\arabic{figure}}
\renewcommand{\theHtable}{S\arabic{table}}
\renewcommand{\theHequation}{S\arabic{equation}}

\section{Scope of this Supplementary Information}\label{sec:supp_scope}

This Supplementary Information provides the technical context underlying Sections~\ref{sec:methods} and~\ref{sec:results} of the main paper. It records the initialization convention, configuration details, mathematical definitions, comparison tables, and additional figures needed for transparency, and shows how simulator-level outputs are distilled into the sweep and factorial summaries reported in the main text. Figures~S1--S5 document the selected-operating-point dynamics and profile-level consequences of adaptive plasticity. Figures~S6--S9 summarize weak-support selection, profile-level support sensitivity, sweep-level context, and cross-family backbone effects. Figures~S10--S12 provide ablation and model-placement views, including the representative recall--order trade-off. Figures~S13--S15 collect factorial plasticity summaries, and Figures~S16--S18 provide the control-family recall, stage-structure, and order-asymmetry overviews. Together, these materials provide the implementation details, secondary quantitative summaries, and supporting figure panels needed to reproduce the reported benchmark.

\section{Supplementary Information for Methods}\label{sec:supp_methods}

\subsection{Stage aggregation and benchmark summaries}\label{sec:supp_stage_aggregation}
The base simulator generated stage-resolved trajectories and mechanistic diagnostics before sweep- or factorial-level aggregation was applied. For any scalar trajectory \(y(k)\), indexed by logged simulation step $k$ and stage $s$, the stage summaries were defined as
\begin{equation}
\bar y_s=\frac{1}{|s|}\sum_{k\in s} y(k),\qquad y_s^{\max}=\max_{k\in s} y(k),\qquad \mathrm{AUC}_s=\operatorname{trapz}\!\bigl(y_s,k_s\bigr).
\end{equation}
Here \(y_s\) denotes the restriction of the trajectory \(y(k)\) to stage \(s\), and \(k_s\) denotes the corresponding vector of logged indices for that stage.
These summaries were then propagated into the weak-support sweep and factorial comparison tables. The resulting AUC values are reported in logged-step units. 

Throughout this Supplementary Information, AUC denotes the time-integrated target trajectory over the specified stage window. For the primary recall metric, larger values indicate that the correct \(A\)-channel remains elevated for longer, or at a stronger sustained level, during the final A-cued recall stage; smaller values indicate weaker, shorter-lived, or less stable recall. The primary recall metric is the final A-cued recall-stage AUC, whereas stage structure and order asymmetry are computed from the same run-level traces after stage segmentation and order-probe construction. 

The stage-structure score is an operational benchmark summary, not a direct biological observable. It was used to quantify whether activity remained aligned with the intended task logic after stage segmentation: past should favor the learned \(A/B\) channels, the novel-exposure stage should favor \(C\), the final A-cued recall stage should favor \(A\), and the intervening rest stages should remain comparatively quiet.

Once the weak-support operating point and control settings were selected on the calibration side of the workflow, they were held fixed for the downstream evaluation. The global settings and perturbation-profile definitions are documented in Tables~\ref{tab:sweep_params} and \ref{tab:profile_params}, while the exact sweep values, matched model-comparison means, and factorial summaries are listed in Tables~\ref{tab:sweep_exact}--\ref{tab:factorial_effects}.

\subsection{Weak-support selection logic and sensitivity}\label{sec:supp_weak_support}
The weak-support sweep ranks only safe candidate values of $\epsilon$ by the normalized score
\begin{equation}
S_{\mathrm{select}}=0.55\,\widetilde{\Delta \mathrm{AUC}}+0.35\,\widetilde{\mathrm{AUC}}_{\mathrm{QL\,full}}+0.10\,\widetilde{O},
\end{equation}
where $\widetilde{\Delta \mathrm{AUC}}$ denotes the normalized full-minus-ablation recall gap, $\widetilde{\mathrm{AUC}}_{\mathrm{QL\,full}}$ denotes the normalized full-model recall score, and $\widetilde{O}$ denotes the normalized order-sensitive summary term. Candidates that alter the no-plasticity quantum-like ablation relative to the zero-floor reference are excluded. In the executed scripts, the corresponding safety check requires agreement with the $\epsilon=0$ ablation reference on mean recall AUC, mean support density, and mean zero-edge fraction. This is why the selected support level should be interpreted as a conservative operating point rather than as a maximal-performance setting. The exact values entering this decision are listed in Table~\ref{tab:sweep_exact}.

Profile-wise sweep maxima were not identical across conditions. In the executed sweep for the full quantum-like model, the baseline and high-noise profiles peaked at $\epsilon=10^{-10}$, the jitter profile peaked at $10^{-8}$, the synapse-loss profile peaked at $0$, and the low-noise profile peaked at $10^{-6}$. The main text therefore treats $\epsilon=10^{-10}$ as a single conservative operating point rather than as a profile-specific optimum. Those profile-wise maxima are useful as a sensitivity analysis, but they are not the basis of the primary benchmark claims. Throughout the factorial analyses, backbone-on denotes the selected weak-support condition with \(\epsilon=10^{-10}\), whereas backbone-off denotes the zero-floor reference with \(\epsilon=0\).

\subsection{Classical-control calibration}\label{sec:supp_control_calibration}
The classical controls were calibrated per profile, support mode, and plasticity combination by searching small two-parameter grids over a stimulus-amplitude scale and a dynamical-rate scale. The executed candidate grids were $(0.70,0.85,1.00,1.15)\times(0.70,0.90,1.10)$ for the real no-phase control and $(0.45,0.60,0.80,1.00)\times(0.60,0.90,1.20)$ for the Markov-rate control. The quantum-like models used the native setting $(1,1)$. The corresponding matched evaluation means are summarized in Table~\ref{tab:model_compare_means}, and the average factorial effects are summarized in Table~\ref{tab:factorial_effects}.

\subsection{Operational benchmark settings}\label{sec:supp_operational_settings}
The cue-mixture coefficients in the main task play distinct roles: $\alpha_{\mathrm{switch}}$ sets the immediate jump at a stage transition, whereas $\alpha_{\mathrm{cont}}$ maintains a weaker sustained cue during the rest of the stage. The executed ratio was fixed before downstream evaluation so that stage boundaries remained clearly expressed without forcing collapse onto the cue template at every logged step. These values should therefore be read as benchmark-operational settings rather than as literature-fitted biological amplitudes.

The same operational logic applies to the adaptive timescale and threshold parameters. The fast and slow update rates were chosen to enforce a clear separation between rapid task-driven change and slower consolidation, with transfer from the fast matrix into the slow matrix remaining subdominant within any single stage. Likewise, the homeostatic target $\theta_{\mathrm{homeo}}$, the heterosynaptic competition term, and the structural create/prune thresholds were fixed as conservative benchmark settings so that the learned network remained neither trivially sparse nor saturated across seeds. These settings are therefore intended to support interpretable comparisons under a frozen benchmark, not to serve as direct fitted estimates of biological rate constants or thresholds. This operational interpretation is conceptually aligned with prior modeling and experimental work distinguishing rapid associative change from slower stabilizing or consolidating processes, including homeostatic regulation, heterosynaptic competition, and structural remodeling \cite{turrigiano2012,zenke2017temporal,toyoizumi2014dynamic,chistiakova2015,holtmaat2009,butz2009activity}.

\subsection{Pipeline ordering and calibration split}\label{sec:supp_pipeline}

The computational workflow was fixed in the following order: first, a sweep wrapper evaluated only the full quantum-like model and the no-plasticity quantum-like ablation across the weak-support grid; second, the selected support value was frozen and written out for reuse; third, the factorial wrapper expanded the frozen benchmark over backbone mode, perturbation profile, model family, and plasticity combination, with the classical controls recalibrated on calibration seeds within each matched condition; and fourth, a collection layer aggregated only disjoint evaluation seeds into model comparisons, backbone comparisons, plasticity-combination contrasts, factorial main effects, and pairwise interactions. Calibration seeds were therefore used only for control tuning, whereas the manuscript-level contrasts were computed on separate evaluation seeds.

\subsection{Mechanistic observables and evidentiary role}\label{sec:supp_mechanistic}
The core simulator generated stage-resolved \(q_A\), \(q_B\), and \(q_C\) trajectories together with mechanistic traces such as \(g_{\max}\). These lower-level observables were compressed into the benchmark summaries reported in the main text. The supplementary mechanistic figures therefore show how the chosen-\(\epsilon\) runs give rise to the paired and aggregated comparisons discussed in the main paper.

More broadly, the emphasis on path dependence and retained sensitivity to initial conditions is consistent with recent work on memory-preserving dynamics and cooperative shielding in complex quantum networks, where interaction structure can preserve information about initial excitation conditions rather than simply maximizing transport or spreading efficiency \cite{ausilio2026memory}. This connection is used only as conceptual context; the present benchmark does not claim a microscopic quantum-network mechanism in neural tissue.

\section{Mathematical details of simulator and aggregation}\label{sec:supp_math_details}

\subsection{Initialization convention}\label{sec:supp_initialization}

For each run seed, the quantum-like state was initialized as a normalized random complex pure state,
\begin{equation}
\psi_0=\frac{\xi+i\eta}{\|\xi+i\eta\|_2},
\end{equation}
where \(\xi,\eta\in\mathbb{R}^{N}\) are independent standard-normal random vectors generated from the run seed. The real-valued no-phase control was initialized from a normalized real random vector,
\begin{equation}
x_0=\frac{\xi}{\|\xi\|_2},
\end{equation}
with \(\xi\in\mathbb{R}^{N}\) generated from the corresponding seed. The Markov-rate control was initialized from a normalized nonnegative random vector,
\begin{equation}
p_{0,i}=\frac{|\xi_i|+10^{-6}}{\sum_j(|\xi_j|+10^{-6})}.
\end{equation}
The learned matrices \(A_{\mathrm{fast}}\), \(A_{\mathrm{slow}}\), and the structural trace were initialized at zero. Thus, the simulations did not begin from a fixed density matrix, a pre-trained associative matrix, or a hand-selected memory state.

\subsection{Operator construction, clipping, and weak-support floor}\label{sec:supp_operator_construction}
Let
\begin{equation}
\operatorname{sym}_0(X)=\tfrac{1}{2}(X+X^\top)-\operatorname{diag}(X),
\end{equation}
and let $\operatorname{clip}_{w_{\max}}(x)=\max(-w_{\max},\min(w_{\max},x))$ act elementwise with $w_{\max}=0.6$. The instantaneous learned matrix before weak-support flooring is
\begin{equation}
A_{\mathrm{plastic}}^{\star}(t)=\operatorname{clip}_{w_{\max}}\!\left(\operatorname{sym}_0\bigl(A_{\mathrm{slow}}(t)+A_{\mathrm{fast}}(t)\bigr)\right).
\end{equation}
For backbone-on runs, the code first builds a sparse template $C$ for the selected topology and then rescales it row-wise to the support budget $\epsilon$ by two rounds of row normalization and resymmetrization,
\begin{equation}
B^{(n+1/2)}_{ij}=\begin{cases}
\epsilon\, B^{(n)}_{ij}/\sum_\ell |B^{(n)}_{i\ell}|, & \sum_\ell |B^{(n)}_{i\ell}|>0,\\
0, & \text{otherwise,}
\end{cases}
\qquad
B^{(n+1)}=\operatorname{sym}_0\!\left(B^{(n+1/2)}\right),
\end{equation}
starting from $B^{(0)}=C$. The floor is then restricted to the nonzero support of $A_{\mathrm{plastic}}^{\star}$, with signs inherited from the learned matrix, and the effective operator becomes
\begin{equation}
A_{\mathrm{eff},ij}(t)=\operatorname{clip}_{w_{\max}}\!\left(\operatorname{sign}\bigl(A_{\mathrm{plastic},ij}^{\star}(t)\bigr)\max\!\left(|A_{\mathrm{plastic},ij}^{\star}(t)|,|B_{ij}|\right)\right).
\end{equation}
This is why the weak-support factor is a floor-only support rule rather than a free additive network: it can stabilize existing nonzero support, but it cannot introduce arbitrary dense connectivity where the learned matrix has no support.

\subsection{Module-mask construction}\label{sec:supp_module_mask}

The main text defines the Hebbian drive using a module mask \(M\). In the executed simulator, \(M\) was obtained by inferring \(K=4\) modules from \(|A_{\mathrm{slow}}|\) through normalized spectral clustering \cite{ng2001spectral,newman2006}. Operationally, this means forming
\begin{equation}
W_{\mathrm{mod}}=|A_{\mathrm{slow}}|,\qquad
D_{\mathrm{mod},ii}=\sum_j W_{\mathrm{mod},ij},\qquad
L_{\mathrm{mod}}=I-D_{\mathrm{mod}}^{-1/2}W_{\mathrm{mod}}D_{\mathrm{mod}}^{-1/2},
\end{equation}
then taking the first \(K\) eigenvectors of \(L_{\mathrm{mod}}\), row-normalizing them, and assigning module labels by \(K\)-means in that spectral embedding.

\subsection{Plasticity update rules}\label{sec:supp_plasticity_rules}

The main text introduces the three adaptive mechanisms conceptually. Here we give the executed update rules and parameter values used in the factorial benchmark. Homeostatic stabilization enforced a row-wise \(L^1\) target through a symmetrized excess-load shrinkage. Defining
\begin{equation}
r_i=\sum_j |A_{\mathrm{slow},ij}|,\qquad
e_i=\max(r_i-\theta_{\mathrm{homeo}},0),\qquad
s_{ij}=\frac{e_i+e_j}{2},
\end{equation}
with \(\theta_{\mathrm{homeo}}=2.4\), the update was
\begin{equation}
A_{\mathrm{slow},ij}\leftarrow A_{\mathrm{slow},ij}
-\eta_{\mathrm{homeo}}\,s_{ij}\,\mathrm{sign}\!\bigl(A_{\mathrm{slow},ij}\bigr),
\qquad
\eta_{\mathrm{homeo}}=2\times10^{-4}.
\end{equation}

Heterosynaptic competition penalized weakly matched fast edges in proportion to node-level Hebbian load. Defining
\begin{equation}
D_i=\frac{1}{N}\sum_j |H_{ij}|,\qquad
L_{ij}=\frac{D_i+D_j}{2},\qquad
m_{ij}=\frac{|H_{ij}|}{\max_{k\ell}|H_{k\ell}|},
\end{equation}
the heterosynaptic update was
\begin{equation}
A_{\mathrm{fast},ij} \leftarrow A_{\mathrm{fast},ij}
-\eta_{\mathrm{hetero}}\,L_{ij}\bigl(1-m_{ij}\bigr)
\,\mathrm{sign}\!\bigl(A_{\mathrm{fast},ij}\bigr),
\qquad
\eta_{\mathrm{hetero}}=0.15.
\end{equation}
This update was applied after the Hebbian fast update and before symmetrization and clipping.

Structural plasticity maintained a trace variable,
\begin{equation}
T_{ij}\leftarrow \lambda T_{ij} + (1-\lambda)|H_{ij}|,
\qquad
\lambda=0.995,
\end{equation}
and acted only on weak slow edges satisfying
\(|A_{\mathrm{slow},ij}|<\theta_{\mathrm{prune}}\), with
\(\theta_{\mathrm{prune}}=0.004\). For those candidate edges,
\begin{equation}
A_{\mathrm{slow},ij}\leftarrow A_{\mathrm{slow},ij}
+\Delta_{\mathrm{struct}}\,\mathrm{sign}\!\bigl(A_{\mathrm{fast},ij}\bigr)
\quad \text{if } T_{ij}\ge \theta_{\mathrm{create}},
\end{equation}
and
\begin{equation}
A_{\mathrm{slow},ij}\leftarrow 0
\quad \text{if } T_{ij}<0.5\,\theta_{\mathrm{create}},
\end{equation}
with \(\theta_{\mathrm{create}}=0.020\) and
\(\Delta_{\mathrm{struct}}=0.004\). These rules define the \(2^3\) factorial grid of homeostatic, heterosynaptic, and structural plasticity used in the frozen benchmark.

\subsection{Pipeline aggregation and stage-resolved summaries}\label{sec:supp_pipeline_aggregation}

The simulation workflow was hierarchical. A base simulator generated stage-resolved state trajectories and mechanistic diagnostics for individual runs, while the sweep and factorial wrappers reused the same core dynamics and aggregated those outputs into per-seed, per-condition, and cross-condition summaries. At the trajectory level, the simulator tracked the evolving effective operator \(A_{\mathrm{eff}}(t)\) together with observables such as \(q_A(t)\), \(q_B(t)\), \(q_C(t)\), spectral-gap summaries, leading-mode occupancy measures, and modularity-style diagnostics.

For any scalar trajectory \(y(k)\) indexed by logged simulation step \(k\) and stage \(s\in\{\mathrm{past},\mathrm{rest}_1,\mathrm{new},\mathrm{rest}_2,\mathrm{recallA}\}\), the stage summaries were
\begin{equation}
\bar y_s=\frac{1}{|s|}\sum_{k\in s} y(k),\qquad
y^{\max}_s=\max_{k\in s} y(k),\qquad
\mathrm{AUC}_s=\sum_{k=k_{\mathrm{start}}}^{k_{\mathrm{end}}-1}\frac{y(k)+y(k+1)}{2}\,\Delta k,
\end{equation}
where \(k_{\mathrm{start}}\) and \(k_{\mathrm{end}}\) are the first and last logged indices of the relevant stage window, and \(\Delta k\) is the spacing between successive logged indices. These AUC values are reported in logged-step units rather than physical time units; because \(\Delta t\) is constant across conditions, paired comparisons are invariant to this global scaling.

The recall metric \(\mathrm{AUC}_{\mathrm{recallA}}\) was derived from the \(q_A(k)\) trajectory during the final A-cued recall stage. Stage structure and order asymmetry were computed from the same run-level outputs after stage segmentation and probe construction. Spectral and modular diagnostics were used as mechanistic checks on the selected operating regime rather than as stand-alone success criteria.

\subsection{Spectral, modular, and support diagnostics}\label{sec:supp_spectral_diagnostics}
If $\lambda_1(t)\le \cdots \le \lambda_N(t)$ are the ordered eigenvalues of $A_{\mathrm{eff}}(t)$ and $v_1(t),\ldots,v_N(t)$ the corresponding eigenvectors, then the simulator computed the extreme-gap diagnostics
\begin{equation}
g_{\min}(t)=\lambda_2(t)-\lambda_1(t),
\qquad
g_{\max}(t)=\lambda_N(t)-\lambda_{N-1}(t),
\end{equation}
as well as the state occupancy of the extreme modes,
\begin{equation}
p_{\mathrm{top}}(t)=|\langle v_N(t),\psi(t)\rangle|^2,
\qquad
p_{\mathrm{bot}}(t)=|\langle v_1(t),\psi(t)\rangle|^2,
\end{equation}
and the top-mode stimulus overlaps,
\begin{equation}
q_A^{\mathrm{top}}(t)=|\langle S_A,v_N(t)\rangle|^2,
\quad
q_B^{\mathrm{top}}(t)=|\langle S_B,v_N(t)\rangle|^2,
\quad
q_C^{\mathrm{top}}(t)=|\langle S_C,v_N(t)\rangle|^2.
\end{equation}

For the inferred labels $c_i$, the simulator also reports mean within-module and between-module slow-weight magnitudes,
\begin{equation}
\overline{W}_{\mathrm{within}}=\operatorname{mean}_{i\neq j,\,c_i=c_j}|A_{\mathrm{slow},ij}|,
\qquad
\overline{W}_{\mathrm{between}}=\operatorname{mean}_{c_i\neq c_j}|A_{\mathrm{slow},ij}|,
\end{equation}
and the weighted modularity of $|A_{\mathrm{slow}}|$,
\begin{equation}
Q=\frac{1}{2m}\sum_{ij}\left(W_{ij}-\frac{k_i k_j}{2m}\right)\delta(c_i,c_j),
\qquad
k_i=\sum_j W_{ij},
\qquad
2m=\sum_i k_i.
\end{equation}
Here \(c_i\) is the inferred module label of node \(i\), and \(\delta(c_i,c_j)\) is the Kronecker delta, equal to \(1\) when nodes \(i\) and \(j\) are assigned to the same module and \(0\) otherwise. These are the quantities behind the modularity-style mechanistic panels.

For any off-diagonal edge set \(E=\{(i,j): i\neq j\}\), weak-edge threshold \(\tau_{\mathrm{weak}}\), and generic counting threshold \(\tau\), the support summaries are
\begin{align}
f_{0}(A)&=\frac{1}{|E|}\sum_{(i,j)\in E}\mathbf{1}\{|A_{ij}|\le 10^{-12}\},\\
f_{\mathrm{weak}}(A)&=\frac{1}{|E|}\sum_{(i,j)\in E}\mathbf{1}\{10^{-12}<|A_{ij}|<\tau_{\mathrm{weak}}\},\\
\rho_{\mathrm{supp}}(A)&=\frac{1}{|E|}\sum_{(i,j)\in E}\mathbf{1}\{|A_{ij}|>10^{-12}\},\\
E_{\tau}(A)&=\sum_{i<j}\mathbf{1}\{|A_{ij}|>\tau\}.
\end{align}
In the executed benchmark, these summaries were used both descriptively and as safety checks in the support-selection stage.

\subsection{Classical-control calibration objective}\label{sec:supp_calibration_objective}
The classical controls were not tuned directly on recall AUC. Instead, calibration minimized a penalty that rewards stage selectivity while discouraging saturated or nearly flat traces. \(\mathrm{stage}_X\) denotes the mean value of channel \(q_X\) over the named stage; for example, \(\mathrm{past}_A\) is the mean \(A\)-channel value during the past stage and \(\mathrm{new}_C\) is the mean \(C\)-channel value during the new stage. Let
\begin{equation}
\begin{split}
S_{\mathrm{cal}}
&=
(\mathrm{past}_A+\mathrm{past}_B-2\,\mathrm{past}_C)\\
&+
(2\,\mathrm{new}_C-\mathrm{new}_A-\mathrm{new}_B)\\
&-
1.5\,(\mathrm{rest1}_{\max}+\mathrm{rest2}_{\max}),
\end{split}
\end{equation}
where
\begin{align*}
\mathrm{rest1}_{\max}
&=\max\{\mathrm{rest1}_A,\mathrm{rest1}_B,\mathrm{rest1}_C\},\\
\mathrm{rest2}_{\max}
&=\max\{\mathrm{rest2}_A,\mathrm{rest2}_B,\mathrm{rest2}_C\}.
\end{align*}
The saturation penalty used in the code is
\begin{equation}
P_{\mathrm{sat}}=\sum_{X\in\{A,B,C\}}\left[2\,\mathbf{1}\!\left(\operatorname{std}(q_X)<10^{-6}\right)+\mathbf{1}\!\left(\max q_X>0.995\right)\right],
\end{equation}
where \(q_X\) denotes the channel-occupancy trajectory for stimulus class \(X\in\{A,B,C\}\) on the calibration run. The quantity minimized on the calibration split is therefore
\begin{equation}
\mathcal{L}_{\mathrm{cal}}=-(S_{\mathrm{cal}}-P_{\mathrm{sat}}).
\end{equation}
This matters for interpretation because the classical baselines were deliberately calibrated toward stage selectivity rather than handicapped by leaving them untuned.

\subsection{Paired contrasts, factorial effects, and effect size}\label{sec:supp_paired_contrasts}

All factorial summaries were computed on matched evaluation cells. For a matched-condition metric \(m\), with \(f\) and \(g\) denoting binary off/on indicators for the plasticity mechanisms under consideration, the factorial main effect of \(f\) was computed as
\begin{equation}
E_f = \overline{m\,|\,f=1} - \overline{m\,|\,f=0},
\end{equation}
where the averages are taken over matched cells while averaging over the other plasticity mechanisms. Pairwise interactions were computed as
\begin{equation}
I_{fg}=\overline{m}_{11}-\overline{m}_{10}-\overline{m}_{01}+\overline{m}_{00},
\end{equation}
where \(\overline{m}_{ab}\) is the mean of \(m\) over matched cells with \(f=a\) and \(g=b\). This decomposition separates first-order plasticity effects from non-additive interactions among homeostatic, heterosynaptic, and structural plasticity. The all-off condition in these factorial summaries refers only to the three additional plasticity mechanisms. It should not be confused with the no-plasticity quantum-like ablation, which disables the adaptive learning process itself and is treated as a separate model family.

For any matched contrast with within-seed difference vector \(d=(d_1,\ldots,d_n)\), the reported paired effect size was
\begin{equation}
d_z=\frac{\overline{d}}{s_d},
\end{equation}
where \(\overline{d}\) and \(s_d\) are the sample mean and sample standard deviation of the paired differences. The factorial main effects and pairwise interactions reported in the paper are therefore averages of matched within-seed changes, not independent-group effects. This same paired logic was used for model-family contrasts, backbone-on versus backbone-off contrasts, and plasticity-combination comparisons. 

For the bootstrap confidence intervals in Table~\ref{tab:principal_seed_ci}, matched differences were first reduced to one value per evaluation seed. For model-family contrasts, the difference for each seed was averaged over the matched backbone, profile, and plasticity-combination cells available for that comparison. For the no-plasticity comparison, the full quantum-like model under the no-extra-plasticity condition was matched to the no-plasticity quantum-like ablation across backbone mode and perturbation profile. For factorial effects, the corresponding within-seed main effect or interaction was computed by averaging over matched condition cells within that seed. The reported intervals are percentile bootstrap \(95\%\) confidence intervals from \(20{,}000\) resamples of the \(30\) evaluation seeds.

\section{Supplementary tables}\label{sec:supp_tables}
The supplementary tables are ordered from configuration disclosure to result-level support. Tables~\ref{tab:code_map} and \ref{tab:param_rationale} translate the internal script labels into manuscript terminology and record the operational rationale for the main executed parameter values. Tables~\ref{tab:sweep_params} and \ref{tab:profile_params} document the global executed settings and perturbation profiles. Table~\ref{tab:sweep_exact} gives the exact weak-support sweep together with the selection score. Tables~\ref{tab:model_compare_means}, \ref{tab:principal_seed_ci}, and \ref{tab:factorial_effects} summarize the principal matched model-comparison, seed-level uncertainty, and factorial results used in the main text, while the remaining supplementary tables retain broader factorial and model-family summaries that underlie the main-paper figures.

\begin{table}[h]
\centering
\caption{Translation of internal script labels into manuscript terminology. The raw labels are retained here only for reproducibility and are avoided in the running text.}
\label{tab:code_map}
\begin{tabular}{ll}
\toprule
Internal label & Manuscript wording \\
\midrule
\texttt{QL full} & Full quantum-like model \\
\texttt{QL no plasticity} & No-plasticity quantum-like ablation \\
\texttt{real\_nophase} & Real-valued no-phase control \\
\texttt{markov\_rate} & Markov-rate control \\
\texttt{module\_sparse} & Sparse modular scaffold \\
\bottomrule
\end{tabular}
\end{table}

\setlength{\LTcapwidth}{\textwidth}
\setlength{\LTleft}{0pt}
\setlength{\LTright}{0pt}
\renewcommand{\arraystretch}{1.15}
\setlength{\tabcolsep}{4pt}

\renewcommand{\arraystretch}{1.15}
\setlength{\extrarowheight}{2pt}
\begin{longtable}{
>{\justifying\setlength{\parindent}{0pt}\arraybackslash}p{0.15\linewidth}
@{\hspace{0.025\linewidth}}
>{\justifying\setlength{\parindent}{0pt}\arraybackslash}p{0.23\linewidth}
@{\hspace{0.025\linewidth}}
>{\justifying\setlength{\parindent}{0pt}\arraybackslash}p{0.32\linewidth}
@{\hspace{0.025\linewidth}}
>{\justifying\setlength{\parindent}{0pt}\arraybackslash}p{0.26\linewidth}
}
\caption{Operational rationale for the main executed parameter values. These entries explain why the reported values were used in the benchmark and how they should be interpreted. Unless noted otherwise, the values are benchmark-operational settings rather than fitted biological constants.}
\label{tab:param_rationale}\\
\toprule
Parameter or setting & Executed value(s) & Why this scale was used here & What future work should test \\
\midrule
\endfirsthead
\toprule
Parameter or setting & Executed value(s) & Why this scale was used here & What future work should test \\
\midrule
\endhead

Cue-onset mixture $\alpha_{\mathrm{switch}}$ &
$0.25$ &
Used to make stage transitions visibly effective without forcing a full template overwrite at every step. &
A broader onset-amplitude sweep and robustness to weaker or stronger cue jumps. \\

Sustained cue mixture $\alpha_{\mathrm{cont}}$ &
$0.04$ &
Kept deliberately smaller than the onset mixture so within-stage dynamics remain shaped by recurrence and plasticity rather than by constant external forcing. &
Alternative onset-to-sustain ratios and profile-specific cue tuning. \\

Fast decay / learning &
\begin{tabular}[t]{@{}l@{}}
$\gamma_{\mathrm{fast}}=2\times10^{-3}$\\
$\eta_{\mathrm{fast}}=0.025$
\end{tabular} &
Places rapid Hebbian-like adaptation on the fast timescale of the benchmark so current task events can change the operator within a stage. &
Sensitivity analysis over faster and slower associative updates. \\

Slow decay / transfer &
\begin{tabular}[t]{@{}l@{}}
$\gamma_{\mathrm{slow}}=2\times10^{-5}$\\
$\kappa=3\times10^{-4}$
\end{tabular} &
Keeps consolidation much slower than the fast component so persistent structure accumulates gradually rather than instantly. &
Systematic fast/slow timescale-ratio sweeps and alternative consolidation rules. \\

Homeostatic target and rate &
\begin{tabular}[t]{@{}l@{}}
$\theta_{\mathrm{homeo}}=2.4$\\
$\eta_{\mathrm{homeo}}=2\times10^{-4}$
\end{tabular} &
Chosen so the slow matrix remains neither trivially sparse nor saturated across seeds, making homeostasis a stabilizing benchmark ingredient rather than a dominant erase-all term. &
Different set-points, adaptive set-points, and profile-dependent stabilization strengths. \\

Heterosynaptic competition rate &
$\eta_{\mathrm{hetero}}=0.15$ &
Large enough to create competition among weakly matched edges, but not so large that it overwhelms the Hebbian fast update on every step. &
Broader competition-strength sweeps and alternative normalization-style rules. \\

Structural trace memory &
$\lambda=0.995$ &
Makes structural updates depend on repeated rather than single-step support, so rewiring remains slower and more discrete than weight plasticity. &
Shorter- and longer-memory structural traces. \\

Structural thresholds and increment &
\begin{tabular}[t]{@{}l@{}}
$\theta_{\mathrm{prune}}=0.004$\\
$\theta_{\mathrm{create}}=0.020$\\
$\Delta_{\mathrm{struct}}=0.004$
\end{tabular} &
Conservative settings that make creation and pruning thresholded events instead of a dense continuous rewrite of the slow matrix. &
Separate creation and pruning sweeps, asymmetric thresholds, and larger rewiring steps. \\

Weak-support sweep grid &
{\scriptsize
\begin{tabular}[t]{@{}l@{}}
$0,10^{-15},10^{-12},$\\
$10^{-10},10^{-9},$\\
$10^{-8},10^{-6}$
\end{tabular}
}&
Covers the zero-floor reference and a broad logarithmic range of weak nonzero support values without presuming monotonicity. &
Finer logarithmic grids near the selected interior regime. \\

Chosen weak support &
$\epsilon=10^{-10}$ &
Selected as a conservative interior operating point because it improved the full model on the screening summaries without rescuing the no-plasticity ablation. &
Alternative conservative-selection rules and profile-specific choices. \\

Control calibration grids &
{\scriptsize
\begin{tabular}[t]{@{}l@{}}
Real no-phase\\
$(0.70,0.85,1.00,1.15)$\\
$\times(0.70,0.90,1.10)$\\[3pt]
Markov-rate\\
$(0.45,0.60,0.80,1.00)$\\
$\times(0.60,0.90,1.20)$
\end{tabular}
}
&
Small grids were used to avoid trivial under-tuning while keeping the comparison computationally tractable and matched across conditions. &
Finer grids, Bayesian search, or multi-objective classical calibration. \\

\bottomrule
\end{longtable}

\begin{table}[h]
\centering
\footnotesize
\caption{Executed sweep and factorial configuration. The exact support grid was used only in the conservative sweep stage; the manuscript text refers to the grid collectively and reserves the exact values for this table.}
\label{tab:sweep_params}
\renewcommand{\arraystretch}{1.08}
\begin{tabularx}{\linewidth}{>{\raggedright\arraybackslash}p{0.34\linewidth} >{\raggedright\arraybackslash}X}
\toprule
Quantity & Value \\
\midrule
Network size $N$ & 90 \\
Stimulus groups & $A=0\!:\!24$, $B=25\!:\!49$, $C=50\!:\!74$ \\
Stage durations & past $=3500$, rest$_1=1200$, new $=2000$, rest$_2=800$, recallA $=1500$ \\
Time step & $\Delta t=0.03$ \\
Stimulus period in past stage & 200 logged steps \\
Stimulus mixing coefficients & $\alpha_{\mathrm{switch}}=0.25$, $\alpha_{\mathrm{cont}}=0.04$ \\
Template leakage & 0.12 \\
Backbone topology & sparse modular scaffold \\
Backbone degree & 4 \\
Module count for spectral mask inference & $K=4$ \\
Sweep grid $\epsilon$ & $0,10^{-15},10^{-12},10^{-10},10^{-9},10^{-8},10^{-6}$ \\
Chosen $\epsilon$ & $10^{-10}$ \\
Sweep seeds & 1--20 \\
Factorial seeds & 1--40 \\
Calibration / evaluation split & 1--10 calibration, 11--40 evaluation \\
$q^\star$ rule & $5/N=0.05556\ldots$ \\
Past-anchor threshold & $0.1\times q_{A,\mathrm{past\ peak}}$ \\
Order probe & 40 logged steps, cue amplitude $=0.10\times$ fitted stimulus-amplitude scale \\
\bottomrule
\end{tabularx}
\end{table}

\begin{table}[h]
\centering
\caption{Executed perturbation profiles. Jitter and synapse-loss perturbations were applied at stage boundaries.}
\label{tab:profile_params}
\begin{tabular}{lccc}
\toprule
Profile & state noise & jitter amplitude & edge-drop probability \\
\midrule
baseline & 0.01 & 0 & 0 \\
low noise & 0.005 & 0 & 0 \\
high noise & 0.02 & 0 & 0 \\
jitter & 0.01 & $10^{-3}$ & 0 \\
synapse loss & 0.01 & $5\times10^{-4}$ & 0.01 \\
\bottomrule
\end{tabular}
\end{table}

\begin{table*}[t]
    \centering
    \caption{Profile-specific weak-support maxima and the chosen common operating point. This table shows that the manuscript uses $\epsilon = 10^{-10}$ as a conservative common operating point rather than as the per-profile optimum in every condition.}
    \label{tab:supp_profilewise_support_maxima}
    \begin{tabular}{lll}
    \toprule
    Profile & Maximum & Role in the manuscript \\
    \midrule
    Baseline & $10^{-10}$ & Matches the chosen common operating point \\
    Low noise & $10^{-6}$ & Sensitivity note only; not used as the main operating point \\
    High noise & $10^{-10}$ & Matches the chosen common operating point \\
    Jitter & $10^{-8}$ & Sensitivity note only; not used as the main operating point \\
    Synapse loss & $0$ & Sensitivity note only; not used as the main operating point \\
    \bottomrule
    \end{tabular}
    \end{table*}

\begin{table}[h]
\centering
\small
\caption{Exact weak-support sweep values used for selection. Columns report the mean recall-stage AUC for the full quantum-like model, the mean recall-stage AUC for the no-plasticity ablation, their mean gap, and the resulting selection score. The selected point maximizes the composite score among safe candidates.}
\label{tab:sweep_exact}
\renewcommand{\arraystretch}{1.15}
\setlength{\tabcolsep}{5pt}
\begin{tabular}{rrrrr}
\toprule
$\epsilon$ & Full & Ablation & Gap & Score \\
\midrule
$0$ & 768.48 & 0.827267 & 767.65 & 0.989952 \\
$10^{-15}$ & 768.48 & 0.827267 & 767.65 & 0.989952 \\
$10^{-12}$ & 768.48 & 0.827267 & 767.65 & 0.989952 \\
$10^{-10}$ & 777.10 & 0.827267 & 776.27 & 1.000000 \\
$10^{-9}$ & 722.48 & 0.827267 & 721.65 & 0.936394 \\
$10^{-8}$ & 737.69 & 0.827267 & 736.86 & 0.954152 \\
$10^{-6}$ & 719.69 & 0.827267 & 718.86 & 0.932948 \\
\bottomrule
\end{tabular}
\end{table}

\begin{table}[h]
\centering
\caption{Evaluation-split aggregated model comparisons for the full quantum-like model averaged over matched evaluation conditions. Positive values favor the full quantum-like model.}
\label{tab:model_compare_means}
\begin{tabular}{lrrrr}
\toprule
Comparator & Mean \(\Delta\)AUC & Mean \(\Delta_{\mathrm{order}}\) & Mean \(\Delta S_{\mathrm{stage}}\) & Fraction \(\Delta\)AUC \(>0\) \\
\midrule
Real-valued no-phase control & 79.1 & 0.629 & 0.982 & 0.55 \\
Markov-rate control & -154.0 & 1.236 & 1.208 & 0.35 \\
No-plasticity quantum-like ablation & 514.5 & -0.587 & -1.972 & 1.00 \\
\bottomrule
\end{tabular}
\end{table}

\begin{table*}[t]
\centering
\footnotesize
\caption{Seed-level bootstrap confidence intervals for principal matched contrasts. Mean differences are computed on evaluation seeds 11--40 after averaging the relevant matched condition cells within each seed. Intervals are percentile bootstrap \(95\%\) confidence intervals from \(20{,}000\) resamples over the \(30\) evaluation seeds. Positive values favor the first model or the named plasticity effect.}
\label{tab:principal_seed_ci}
\renewcommand{\arraystretch}{1.15}
\begin{tabular}{llrrr}
\toprule
Contrast or effect & Metric & Mean difference & 95\% bootstrap CI & \(n\) seeds \\
\midrule
QL full -- Markov-rate & Recall AUC & \(-154.0\) & \([-170.0,\,-137.7]\) & 30 \\
QL full -- Markov-rate & Order asymmetry & \(1.236\) & \([1.220,\,1.252]\) & 30 \\
QL full -- Markov-rate & Stage structure & \(1.208\) & \([1.188,\,1.229]\) & 30 \\
\addlinespace
QL full -- real-valued no-phase & Recall AUC & \(79.1\) & \([46.0,\,111.9]\) & 30 \\
QL full -- real-valued no-phase & Order asymmetry & \(0.629\) & \([0.610,\,0.650]\) & 30 \\
QL full -- real-valued no-phase & Stage structure & \(0.982\) & \([0.895,\,1.077]\) & 30 \\
\addlinespace
Markov-rate -- real-valued no-phase & Recall AUC & \(233.1\) & \([204.6,\,261.8]\) & 30 \\
Markov-rate -- real-valued no-phase & Order asymmetry & \(-0.607\) & \([-0.621,\,-0.592]\) & 30 \\
Markov-rate -- real-valued no-phase & Stage structure & \(-0.226\) & \([-0.316,\,-0.127]\) & 30 \\
\addlinespace
QL full, no-extra-plasticity -- QL no-plasticity ablation & Recall AUC & \(514.5\) & \([441.6,\,590.5]\) & 30 \\
QL full, no-extra-plasticity -- QL no-plasticity ablation & Order asymmetry & \(-0.587\) & \([-0.619,\,-0.553]\) & 30 \\
QL full, no-extra-plasticity -- QL no-plasticity ablation & Stage structure & \(-1.972\) & \([-2.010,\,-1.931]\) & 30 \\
\addlinespace
Homeostatic main effect & Recall AUC & \(157.6\) & \([122.0,\,194.1]\) & 30 \\
Heterosynaptic main effect & Recall AUC & \(-20.6\) & \([-50.5,\,8.1]\) & 30 \\
Structural main effect & Recall AUC & \(36.9\) & \([3.4,\,70.6]\) & 30 \\
Heterosynaptic \(\times\) structural interaction & Recall AUC & \(258.9\) & \([191.3,\,329.1]\) & 30 \\
\bottomrule
\end{tabular}
\end{table*}

\begin{table}[h]
\centering
\caption{Average first-order and pairwise factorial effects on the full quantum-like model final A-cued recall-stage area under the curve across evaluation conditions.}
\label{tab:factorial_effects}
\begin{tabular}{lrr}
\toprule
Factor & mean main effect & mean paired $d_z$ \\
\midrule
heterosynaptic & -20.6 & -0.083 \\
homeostatic & 157.6 & 1.095 \\
structural & 36.9 & 0.172 \\
\midrule
Pair & mean interaction & mean paired $d_z$ \\
\midrule
heterosynaptic + structural & 258.9 & 0.747 \\
homeostatic + heterosynaptic & -57.9 & -0.110 \\
homeostatic + structural & 108.6 & 0.483 \\
\bottomrule
\end{tabular}
\end{table}

\begin{table}[h]
\centering
\caption{Executed evaluation-split factorial coverage by model family. The classical controls were run across the full $2^3$ plasticity grid in both backbone modes; only the quantum-like no-plasticity ablation is restricted to the none condition.}
\label{tab:recovered_factorial_coverage}
\begin{tabular}{lccccc}
\toprule
Model family & backbone modes & plasticity types & profiles & summary rows & note \\
\midrule
QL full & 2 & 8 & 5 & 80 & full $2^3$ grid \\
Real-valued no-phase & 2 & 8 & 5 & 80 & full $2^3$ grid \\
Markov rate & 2 & 8 & 5 & 80 & full $2^3$ grid \\
QL no-plasticity & 2 & 1 & 5 & 10 & none only \\
\bottomrule
\end{tabular}
\end{table}

\begin{table}[h]
\centering
\footnotesize
\caption{Backbone-on minus backbone-off differences averaged across evaluation conditions. Means are averaged over the available profile--combination cells within each model family, and the maximum-absolute columns report the largest shift observed in any matched evaluation cell.}
\label{tab:recovered_backbone_effects}
\begin{tabular*}{\textwidth}{@{\extracolsep{\fill}}lrrrrrr}
\toprule
Model family & \makecell{Mean\\\(\Delta\)AUC} & \makecell{Mean\\\(\Delta_{\mathrm{order}}\)} & \makecell{Mean\\\(\Delta S_{\mathrm{stage}}\)} & \makecell{Max\\\(|\Delta\mathrm{AUC}|\)} & \makecell{Max\\\(|\Delta_{\mathrm{order}}|\)} & \makecell{Max\\\(|\Delta S_{\mathrm{stage}}|\)} \\
\midrule
QL full & -1.218 & -9.52e-05 & 8.36e-04 & 30.14 & 2.33e-03 & 0.0203 \\
Real-valued no-phase & -0.3502 & 3.31e-04 & 9.34e-04 & 8.502 & 6.14e-03 & 0.0118 \\
Markov rate & 1.15e-12 & -7.52e-20 & -1.79e-19 & 2.97e-11 & 8.10e-19 & 1.85e-18 \\
QL no-plasticity & 0.00e+00 & 0.00e+00 & 0.00e+00 & 0.00e+00 & 0.00e+00 & 0.00e+00 \\
\bottomrule
\end{tabular*}
\end{table}

\begin{table}[p]
\centering
\caption{Profile-averaged backbone-on factorial main effects for the three plasticity-enabled model families. Each entry is the average first-order effect of switching the named mechanism on rather than off while averaging over the other two mechanisms.}
\label{tab:recovered_main_effects}
\begin{tabular}{llrrr}
\toprule
Model family & metric & Homeostatic & Heterosynaptic & Structural \\
\midrule
QL full & Recall AUC & 157.6 & -21.79 & 35.63 \\
QL full & Order asymmetry & -9.24e-03 & 0.5652 & 0.0502 \\
QL full & Stage structure & -0.0275 & 1.889 & -0.2519 \\
Real-valued no-phase & Recall AUC & -58.75 & -21.46 & -5.250 \\
Real-valued no-phase & Order asymmetry & 0.0598 & 0.3921 & -0.0326 \\
Real-valued no-phase & Stage structure & 0.1425 & 0.2374 & -0.2971 \\
Markov rate & Recall AUC & 0.00e+00 & -380.7 & 77.27 \\
Markov rate & Order asymmetry & 0.00e+00 & 3.36e-04 & 2.62e-04 \\
Markov rate & Stage structure & 0.00e+00 & 4.04e-04 & 2.43e-04 \\
\bottomrule
\end{tabular}
\end{table}

\begin{table}[!htbp]
\centering
\footnotesize
\caption{Plasticity combination achieving the highest profile-averaged value for each benchmark metric within each model family. Values in parentheses are the corresponding profile-averaged metric means.}
\label{tab:recovered_best_combo_by_metric}
\begin{tabularx}{\textwidth}{@{}l
>{\centering\arraybackslash}X
>{\centering\arraybackslash}X
>{\centering\arraybackslash}X@{}}
\toprule
Model family &
Combination maximizing recall AUC &
Combination maximizing order asymmetry &
Combination maximizing stage structure \\
\midrule
QL full & all three (652.0) & all three (1.563) & homeostatic + heterosynaptic (2.370) \\
Real-valued no-phase & none (514.3) & all three (0.9395) & homeostatic (0.5072) \\
Markov rate & homeostatic + structural (926.1) & all three (0.0278) & all three (-0.0101) \\
\bottomrule
\end{tabularx}
\end{table}

\begin{table}[!htbp]
\centering
\footnotesize
\caption{Backbone-on best-AUC representative cells used in Figure~\ref{fig:supp_order_auc_tradeoff}. For each model--profile pair, the table lists the plasticity combination that maximizes recall AUC and the corresponding coordinates in the plotted recall-AUC--order-asymmetry plane. Plasticity types are abbreviated as H (homeostatic), He (heterosynaptic), and S (structural).}
\label{tab:best_auc_tradeoff_points}
\setlength{\tabcolsep}{3pt}
\renewcommand{\arraystretch}{1.08}
\begin{tabularx}{\textwidth}{@{}l
>{\centering\arraybackslash}X r >{\centering\arraybackslash}p{0.11\textwidth}
>{\centering\arraybackslash}X r >{\centering\arraybackslash}p{0.11\textwidth}
>{\centering\arraybackslash}X r >{\centering\arraybackslash}p{0.11\textwidth}@{}}
\toprule
& \multicolumn{3}{c}{Quantum-like model with plasticity}
& \multicolumn{3}{c}{Real-valued no-phase}
& \multicolumn{3}{c}{Markov rate} \\
\cmidrule(lr){2-4}\cmidrule(lr){5-7}\cmidrule(lr){8-10}
Profile
& Type(s) & AUC & \makecell{Order\\asymmetry}
& Type(s) & AUC & \makecell{Order\\asymmetry}
& Type(s) & AUC & \makecell{Order\\asymmetry} \\
\midrule
Baseline      & H+S  & 641.32  & 1.10787 & S    & 451.88 & 0.28497 & H+S  & 921.35  & 0.027633 \\
Low noise     & H    & 546.67  & 0.79402 & S    & 449.45 & 0.31600 & H+S  & 1048.31 & 0.027633 \\
High noise    & H    & 1049.52 & 1.18292 & He+S & 536.47 & 1.23533 & H+S  & 818.97  & 0.027634 \\
Jitter        & He+S & 621.13  & 1.55217 & S    & 453.75 & 0.28451 & H+S  & 921.29  & 0.027633 \\
Synapse loss  & none & 756.88  & 0.61635 & none & 854.20 & 0.17422 & H+S  & 920.71 & 0.027633 \\
\bottomrule
\end{tabularx}
\end{table}

\clearpage

\section{Supplementary results notes}\label{sec:supp_results_notes}

\subsection{Additional quantitative details for support selection}
The main text reports the conservative logic of the weak-support selection step and retains only the most central aggregate comparison. For completeness, the aggregate full-model recall AUC rose from $768.48$ at $\epsilon=0$ to $777.10$ at $\epsilon=10^{-10}$, then fell again at larger floor values; across the same sweep, the no-plasticity quantum-like ablation remained fixed at $\mathrm{AUC}_{\mathrm{recallA}}=0.827267$ within the summaries used for the safety screen. In the frozen factorial benchmark, toggling the weak structural floor changed the aggregate summaries only slightly for the full quantum-like model, with mean shifts of $\Delta \mathrm{AUC}_{\mathrm{recallA}}=-1.22$, $\Delta_{\mathrm{order}}=-9.5\times10^{-5}$, and $\Delta S_{\mathrm{stage}}=+8.4\times10^{-4}$. These values reinforce the interpretation that the chosen floor acts as a conservative operating constraint rather than as a hidden memory rescue mechanism.

\subsection{Additional mechanistic notes at the chosen operating point}
At $\epsilon=10^{-10}$, the profile-resolved full quantum-like model achieved mean recall AUCs of $694.6$, $593.8$, $990.5$, $639.2$, and $967.4$ across the baseline, low-noise, high-noise, jitter, and synapse-loss profiles, respectively, whereas the no-plasticity quantum-like ablation remained at $0.827$ in every profile. In the corresponding trajectory panels, the full model shows a sustained rise in $q_A(k)$ during the final A-cued recall stage, whereas the ablation remains comparatively flat. The accompanying $g_{\max}$ traces are useful mechanistically because they suggest that the selected-support regime preserves internal spectral organization without the large support-driven distortions that would be expected if the scaffold were dominating the dynamics.

\subsection{Additional matched-comparison and stress-test notes}
For the matched model comparisons, the mean differences are as follows. Relative to the Markov-rate control, the full quantum-like model had mean $\Delta \mathrm{AUC}=-154.0$, mean \(\Delta_{\mathrm{order}}=+1.236\), and mean \(\Delta S_{\mathrm{stage}}=+1.208\). Relative to the real-valued no-phase control, the corresponding means were $+79.1$, $+0.629$, and $+0.982$. Against the Markov-rate control, the quantum-like model was ahead in only $35\%$ of matched evaluation cells on raw recall AUC, but in $100\%$ on order asymmetry and $95\%$ on stage structure; against the real-valued no-phase control, the corresponding positive fractions were $55\%$, $97.5\%$, and $82.5\%$. In the direct comparison between the two classical controls, the Markov-rate control exceeded the real-valued no-phase control in (82.5\%) of matched cells on recall AUC, but only in (0\%) on order asymmetry and (30\%) on stage structure. Relative to the no-plasticity quantum-like ablation, the full quantum-like model also showed a large increase in thresholded recall, with mean \(\Delta\) fraction hit \(q^\star=+0.807\), where \(q^\star=5/N\) is defined in Table~\ref{tab:sweep_params}.

The high-noise profile is the clearest stress-test condition in the present benchmark. Under weak support, the mean paired recall advantage of the quantum-like model over the Markov-rate control became positive in that profile ($+191.5$ AUC units on average), while the order-asymmetry and stage-structure advantages remained strongly positive. Because the same $2^3$ plasticity grid was also executed for the classical controls, we additionally checked whether the matched quantum-like trade-off could be reproduced simply by retuning plasticity within those families. For the Markov-rate family, profile-averaged recall AUC spanned $468.2$ to $926.1$ across plasticity combinations, yet profile-averaged order asymmetry remained confined to $0.0272$--$0.0278$ and profile-averaged stage-structure scores remained negative near $-0.011$. For the real-valued no-phase family, the none condition gave the highest profile-averaged recall AUC ($514.3$), whereas the all-three combination gave the largest order asymmetry ($0.939$) and homeostatic plasticity gave the largest positive stage-structure score ($0.507$). These broader numerical summaries clarify why the classical families do not collapse onto the same joint benchmark profile as the full quantum-like model.

\section{Supplementary figures}\label{sec:supp_figures}

\subsection{Figure S1: Combined all-profile overlay dashboard}
\begin{figure}[t]
\centering
\includegraphics[width=0.99\linewidth]{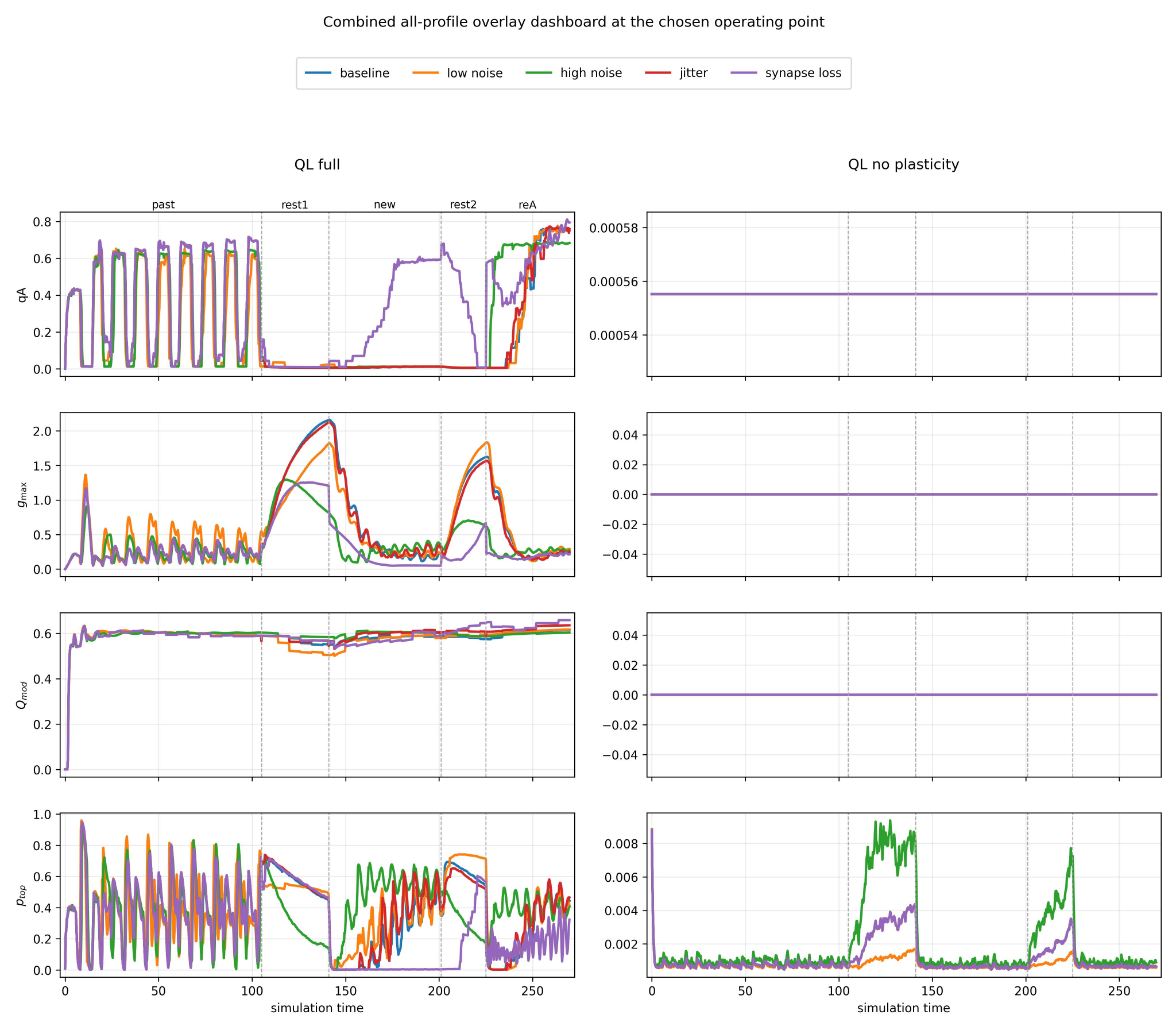}
\caption{Combined all-profile overlay dashboard at the selected weak-support operating point. Rows show \(q_A\), \(g_{\max}\), \(Q_{\mathrm{mod}}\), and \(p_{\mathrm{top}}\); columns compare the full quantum-like model with the no-plasticity ablation. Dashed vertical lines mark stage boundaries.}
\label{fig:supp_overlay_dashboard}
\end{figure}

Figure~\ref{fig:supp_overlay_dashboard} summarizes the chosen-support run across profiles and observables. The left column shows the full quantum-like model and the right column shows the no-plasticity ablation. Rows track four observables: \(q_A\) (recall-channel activation), \(g_{\max}\) (a spectral-gap diagnostic), \(Q_{\mathrm{mod}}\) (a modularity-like segregation summary), and \(p_{\mathrm{top}}\) (dominant-mode occupancy). The dashed vertical lines mark stage boundaries. 

The full model shows profile-dependent stage structure across the trajectory-level observables. In particular, \(q_A\) is activated during the past stage, suppressed during the intervening stages, and rebuilt during final A-cued recall, whereas the no-plasticity ablation remains close to floor. The synapse-loss profile begins to separate earlier than the other profiles, while the high-noise profile rises most strongly by the end of recallA. In the no-plasticity column, by contrast, \(q_A\) is essentially flat at a near-zero baseline across all stages. The key point is therefore not simply that the full model is larger, but that the full model preserves stage selectivity while still producing substantial late recall. The fluctuations in the full-model traces reflect the time-resolved imprint of the staged driving protocol under different perturbation profiles.

The \(g_{\max}\) row explains why the chosen run is still dynamically structured. In the full model, the spectral-gap signal rises sharply around the driven transitions and then relaxes between them, with the largest peaks appearing near the new and recallA stages. The high-noise and synapse-loss profiles suppress this signal more strongly than baseline or low noise, but do not erase it. In the no-plasticity ablation, \(g_{\max}\) is essentially pinned near zero throughout. This indicates that the selected-support full model preserves nontrivial internal organization, whereas the ablation does not.

The \(Q_{\mathrm{mod}}\) row is comparatively stable, which is itself informative. In the full model it remains near \(0.58\)--\(0.64\) across all stages and profiles, indicating that the selected operating point preserves a persistent modular separation rather than creating a transient burst of organization that appears only during recall. The profile fluctuations are modest and interpretable: low noise dips somewhat in the new stage and synapse loss climbs slightly by recallA. The ablation again stays effectively at zero on the plotting scale, indicating an absence of the same evolving modular support structure.

Finally, the \(p_{\mathrm{top}}\) row shows how strongly the leading mode is occupied. In the full model, \(p_{\mathrm{top}}\) fluctuates strongly across stages and profiles, especially under high noise, low noise, and jitter, which indicates active redistribution of dominant-mode occupancy rather than a static dominant state. In the no-plasticity ablation, the signal is much smaller and more weakly structured, with high noise again the clearest exception. Conjointly, the four rows support the main-paper claim that the selected-support full model remains mechanically differentiated across profiles, whereas the no-plasticity ablation does not merely recall less; it also fails to build the same layered dynamical structure.

\subsection{Figure S2: Compactness, density, and plastic-strength summary}
\begin{figure}[t]
\centering
\includegraphics[width=0.98\linewidth]{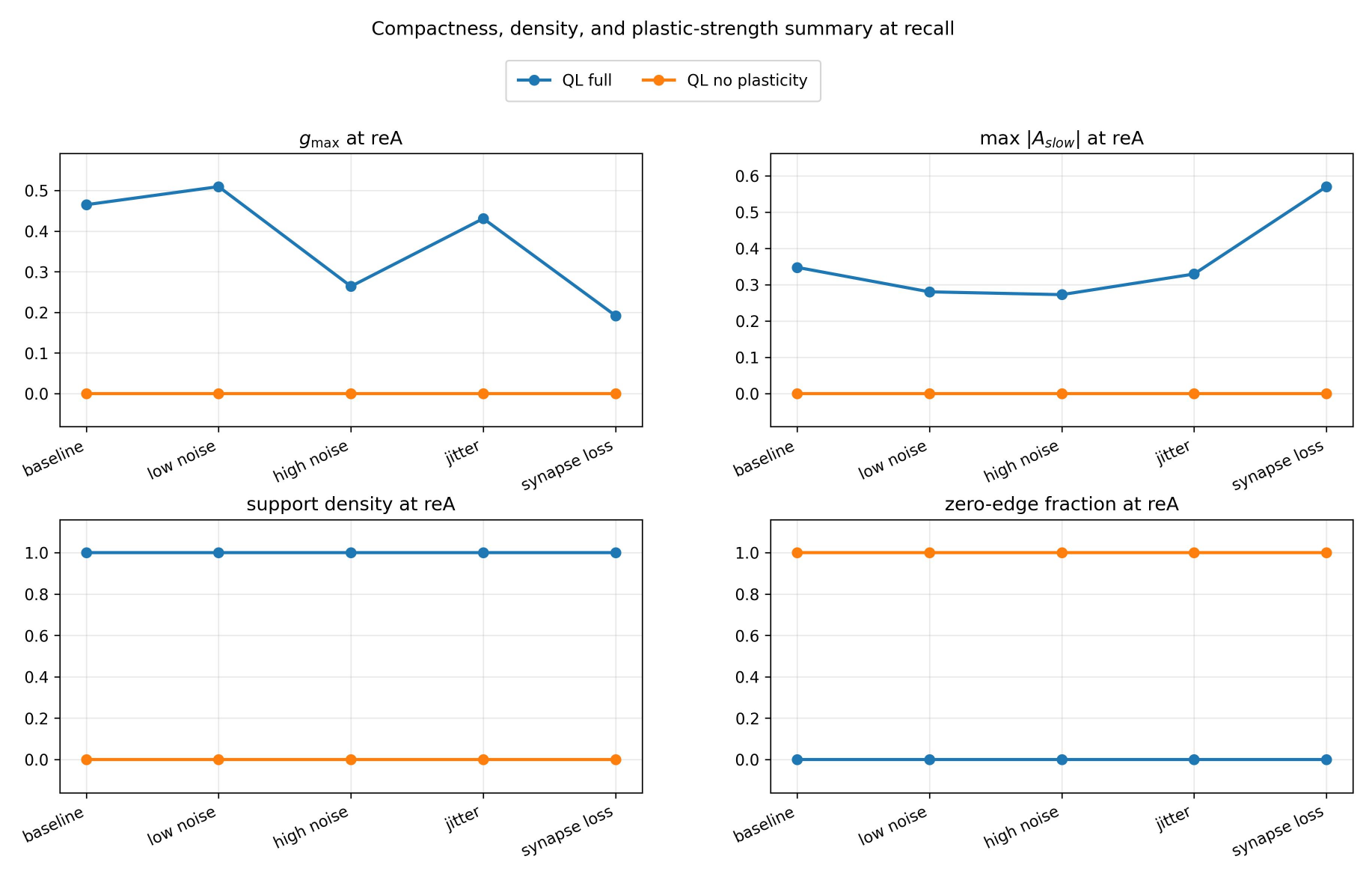}
\caption{Compactness, density, and plastic-strength summary at recallA for the selected weak-support run. The panels report recall-stage \(g_{\max}\), \(\max|A_{\mathrm{slow}}|\), support density, and zero-edge fraction for the full quantum-like model and its no-plasticity ablation.}
\label{fig:supp_network_compactness}
\end{figure}

Figure~\ref{fig:supp_network_compactness} compresses the chosen-support run further by showing only recall-stage summaries. The four panels summarize network-level properties. \(g_{\max}\) at recallA indicates whether a substantial spectral separation is still present late in the trial. \(\max|A_{\mathrm{slow}}|\) at recallA summarizes how strong the learned slow matrix becomes under each profile. Support density and zero-edge fraction show whether the learned network remains fully occupied or collapses toward an empty graph. Across all four panels the contrast is pronounced: the full model retains nonzero spectral organization, nonzero learned weight magnitude, full support density, and zero zero-edge fraction, whereas the no-plasticity ablation remains exactly at the opposite structural limit.

This panel translates network involvement into concrete profile-level summaries. The full model does not only recall more; it also ends recall with a denser, stronger, and spectrally more differentiated network state. Synapse loss is especially informative because it still preserves support density while showing the largest \(\max|A_{\mathrm{slow}}|\), which implies that the full model compensates by strengthening the slow matrix rather than by abandoning structure.

\subsection{Figure S3: Stage-resolved summary lines}
\begin{figure}[t]
\centering
\includegraphics[width=0.99\linewidth]{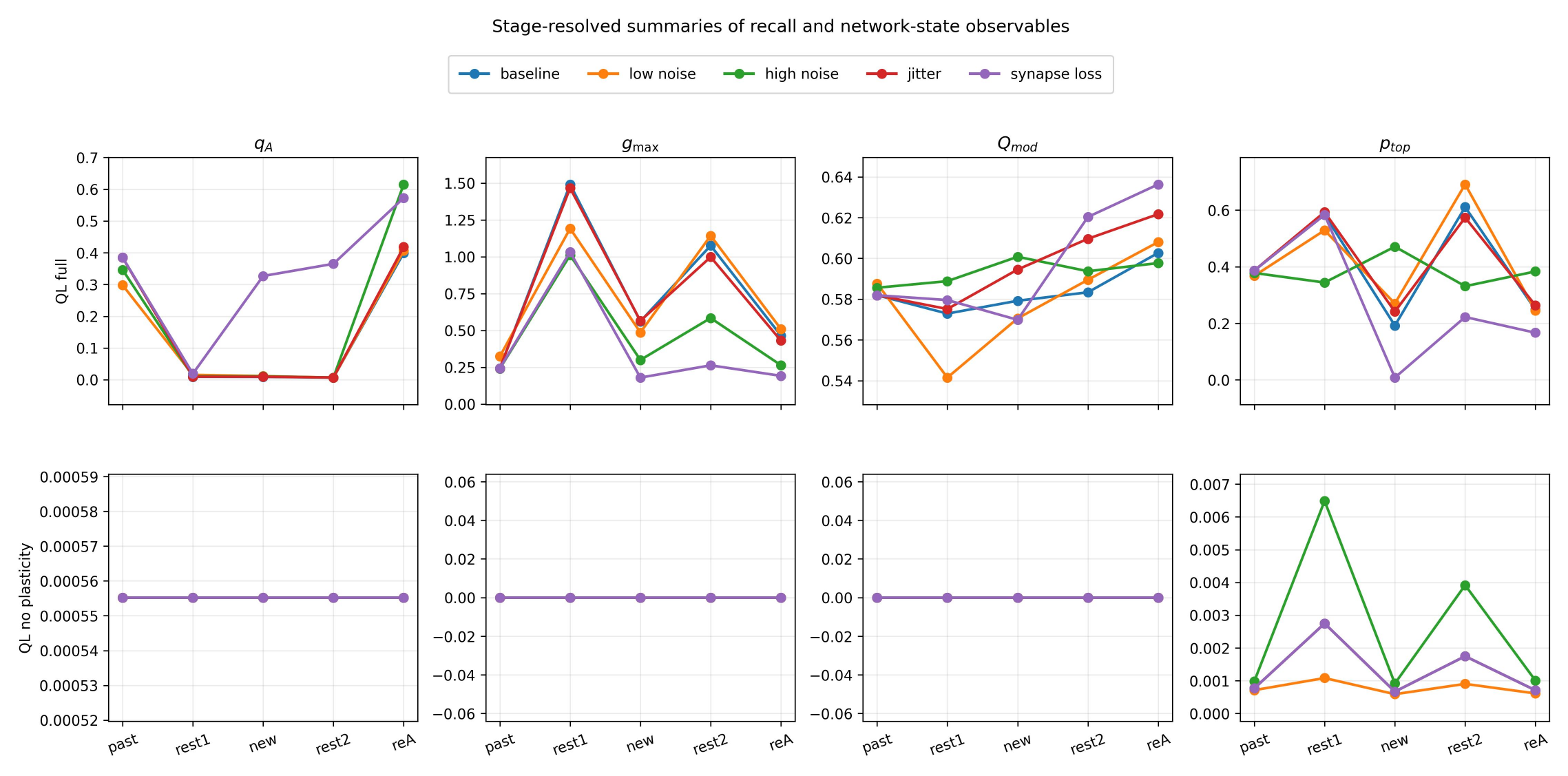}
\caption{Stage-resolved means of \(q_A\), \(g_{\max}\), \(Q_{\mathrm{mod}}\), and \(p_{\mathrm{top}}\) for the chosen-support run. The top row reports the full quantum-like model across the five perturbation profiles; the bottom row reports the no-plasticity ablation on the same observables.}
\label{fig:supp_stage_resolved}
\end{figure}

Figure~\ref{fig:supp_stage_resolved} reduces the full trajectories to stage means. This view asks whether profile differences arise from one anomalous stage or from systematic changes in stage structure across the protocol. In the top row, the full model shows profile-specific but interpretable stage means: \(q_A\) is high in the past stage, near-suppressed in the middle of the protocol, and rises again in recallA; \(g_{\max}\) shows stage-dependent peaks rather than remaining flat; \(Q_{\mathrm{mod}}\) stays in a narrow positive band; and \(p_{\mathrm{top}}\) redistributes across stages. The bottom row shows that the ablation does not maintain the same stage-resolved structure. Its \(q_A\), \(g_{\max}\), and \(Q_{\mathrm{mod}}\) summaries stay close to their respective floors, while \(p_{\mathrm{top}}\) retains only a weak residual profile dependence.

Relative to Figure~\ref{fig:supp_overlay_dashboard}, this figure emphasizes interpretability. The overlay dashboard preserves detailed fluctuations, whereas the stage-resolved means show which stage carries which model difference. High noise retains low mid-protocol \(q_A\) but produces the largest recall-stage \(q_A\), while also suppressing \(g_{\max}\) more strongly than the other profiles. Synapse loss shows the strongest late rise in \(Q_{\mathrm{mod}}\).

\subsection{Figure S4: Profile-wise deltas between the full model and the no-plasticity ablation}
\begin{figure}[t]
\centering
\includegraphics[width=0.98\linewidth]{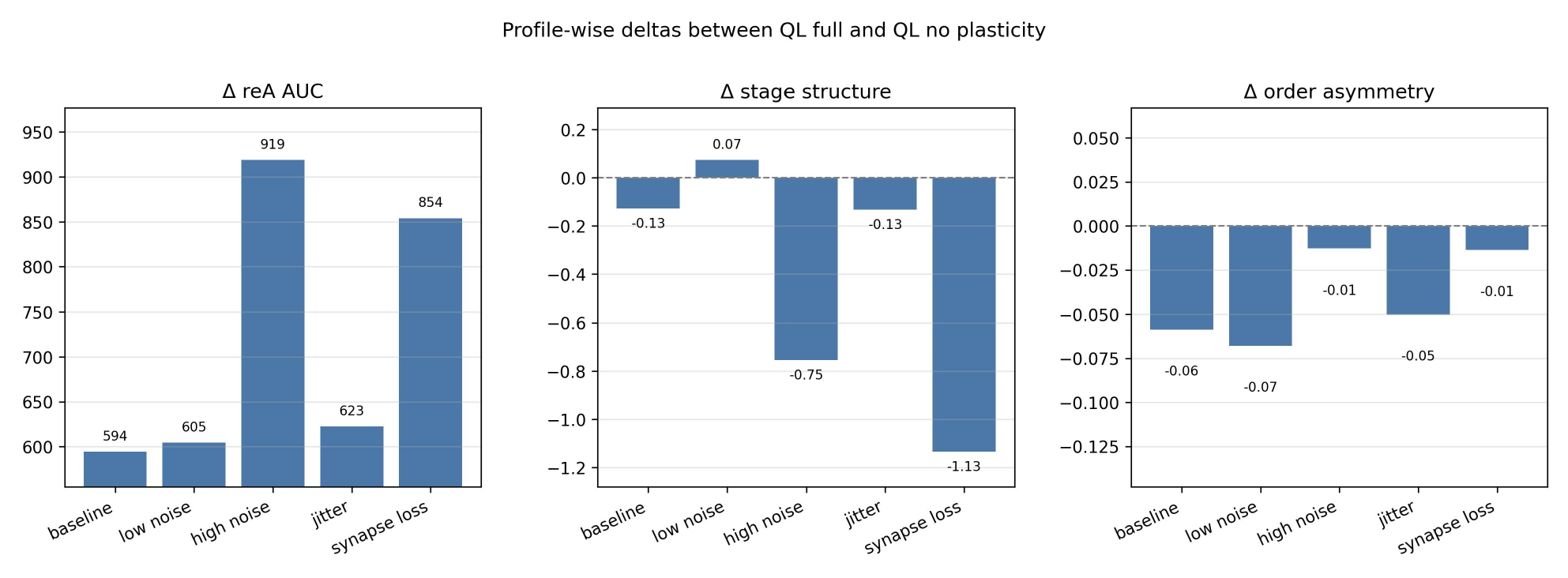}
\caption{Profile-resolved differences between the full quantum-like model and its no-plasticity ablation at the selected weak-support operating point. The panels show changes in recall AUC, stage structure, and order asymmetry.}
\label{fig:supp_model_deltas}
\end{figure}

Figure~\ref{fig:supp_model_deltas} summarizes the comparison in the simplest profile-wise form. The first panel shows that the full model has a large recall advantage over the ablation in every profile, with the largest gains in high noise and synapse loss. The second and third panels show something more subtle: those recall gains are not accompanied by uniformly positive changes in stage structure or order asymmetry. In fact, the full model is slightly worse on those auxiliary metrics in most profiles relative to the ablation.

This pattern helps separate stage structure from recall performance. The ablation can preserve a structured but nearly non-recalling state, whereas the full model trades some raw stage organization and order-asymmetry magnitude for successful recall. 

\subsection{Figure S5: Cross-profile metric summary at the chosen operating point}
\begin{figure}[t]
\centering
\includegraphics[width=0.99\linewidth]{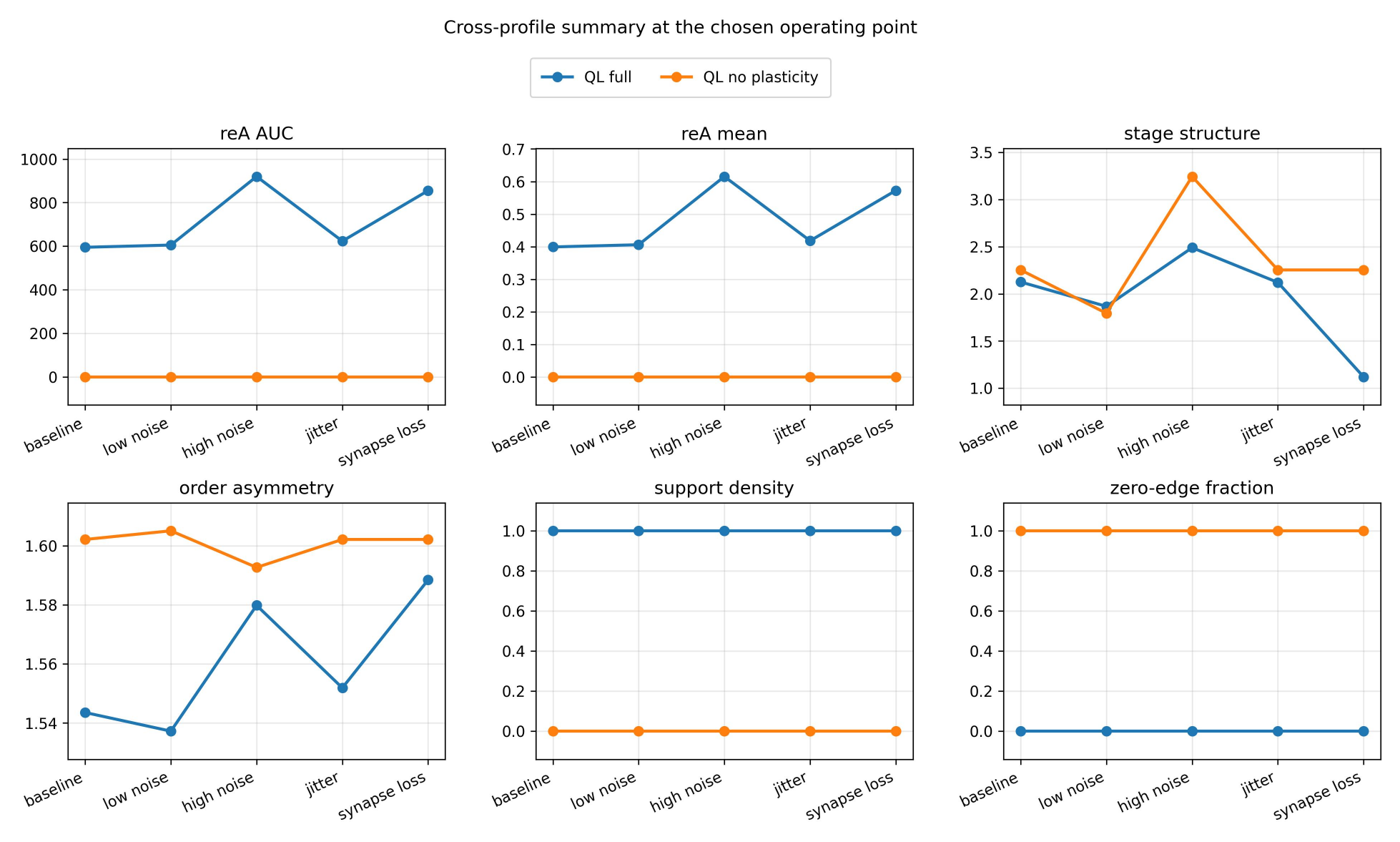}
\caption{Cross-profile summary at the chosen weak-support operating point for the full quantum-like model and the no-plasticity ablation. The panels report recall AUC, recall mean, stage structure, order asymmetry, support density, and zero-edge fraction.}
\label{fig:supp_profile_metric_summary}
\end{figure}

Figure~\ref{fig:supp_profile_metric_summary} gathers six profile-level summaries in a single compact panel set: recall AUC, recall mean, stage structure, order asymmetry, support density, and zero-edge fraction. It sits conceptually between Figures~\ref{fig:supp_model_deltas} and \ref{fig:supp_overlay_dashboard}. Relative to the model-delta figure, it shows absolute values rather than only differences. Relative to the dashboard, it removes the time dimension and keeps only profile-level endpoints. The full model is higher on recall AUC and recall mean in every profile, while the ablation remains effectively at floor. Stage structure and order asymmetry, by contrast, are not uniformly higher in the full model. Support density and zero-edge fraction show the clearest structural contrast: the full model remains fully occupied, whereas the ablation remains effectively empty.

Altogether, Figs.~S1--S5 support four points about the chosen operating point. First, the selected-support full model remains stage-selective and profile-differentiated at the trajectory level.
Second, the same pattern is retained after aggregation into compact recall-stage, stage-mean, and cross-profile summaries. Third, the full-minus-ablation comparison is strongly positive for recall while remaining mixed on auxiliary metrics, which increases the credibility of the benchmark. Fourth, the chosen-support regime therefore has a mechanistic signature that is consistent with the main-paper reading of weak support as a stabilizer of a plasticity-driven operating point.

Figures~S6--S10 extend this view to weak-support selection, profile-level support sensitivity, sweep-level context, cross-family backbone effects, and the no-extra-plasticity ablation gap.

\subsection{Figure S6: Scalar support-selection traces}

\begin{figure}[t]
\centering
\includegraphics[width=0.95\linewidth]{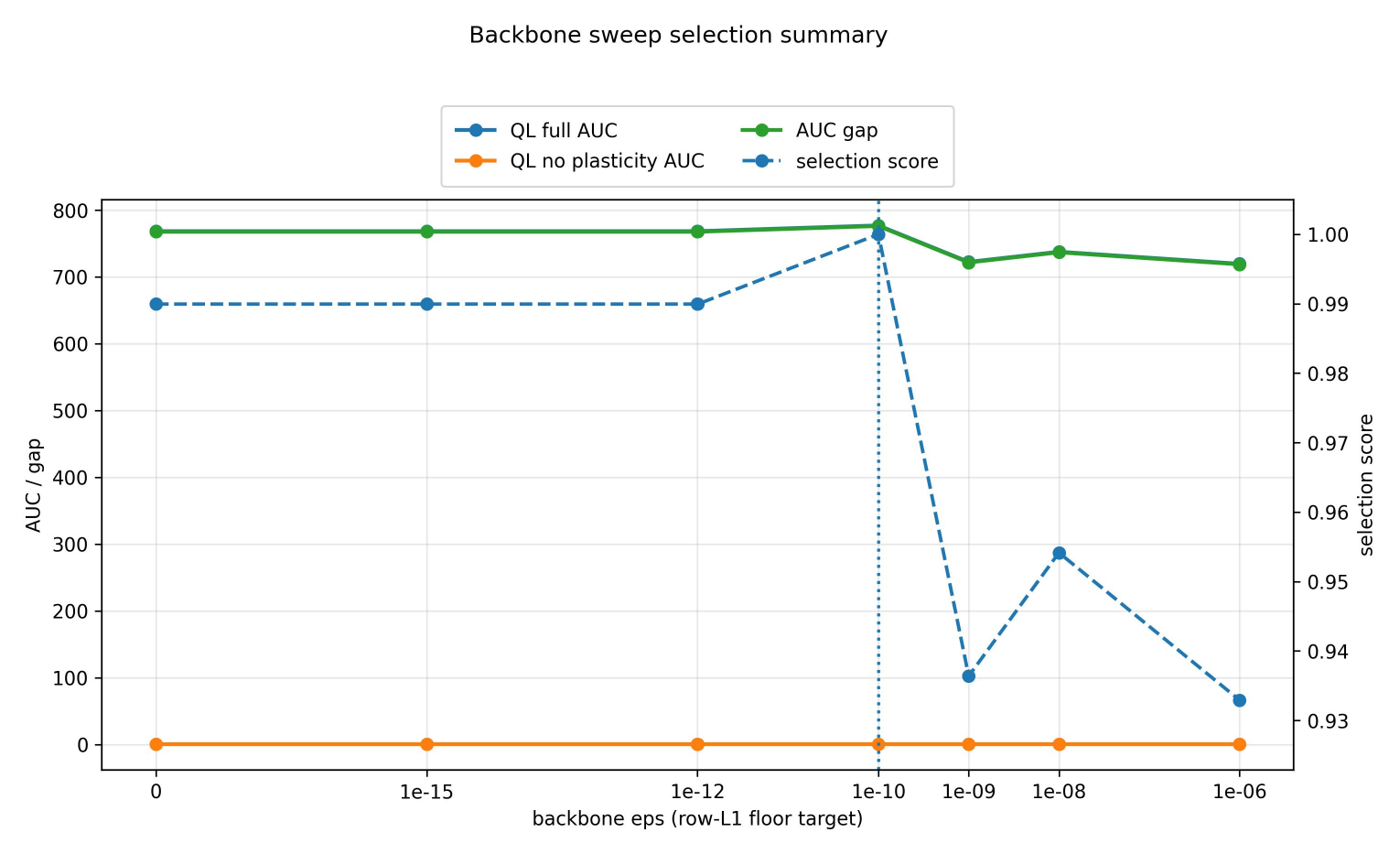}
\caption{Scalar support-selection traces across the weak-support sweep. The curves report the full-model recall AUC, the no-plasticity recall AUC, the full-minus-ablation AUC gap, and the composite selection score used to choose the conservative operating point. The key visual feature is the shared interior maximum at $\epsilon = 10^{-10}$ together with the invariance of the no-plasticity ablation curve.}
\label{fig:supp_sweep_scalar}
\end{figure}

Figure~\ref{fig:supp_sweep_scalar} gives a compact scalar view of the weak-support selection step. The full-model recall AUC improves to an interior maximum, the no-plasticity ablation remains fixed, the AUC gap peaks at the same interior point, and the composite score therefore selects \(\epsilon=10^{-10}\). The panel complements the profile-resolved sweep summaries by showing the conservative selection rule in a single view.

\subsection{Figure S7: Profile-specific support sensitivity}
\begin{figure}[t]
\centering
\includegraphics[width=0.98\linewidth]{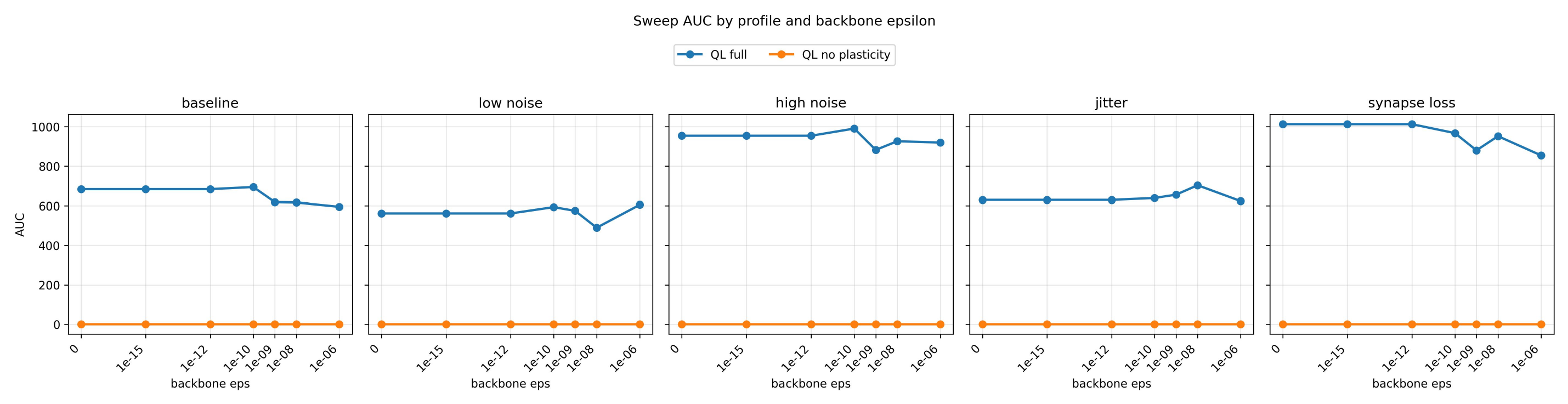}
\caption{Profile-wise recall AUC across the weak-support sweep for the full quantum-like model and the no-plasticity ablation. Each panel corresponds to one perturbation profile.}
\label{fig:supp_sweep_profile_lines}
\end{figure}

Figure~\ref{fig:supp_sweep_profile_lines} shows why the chosen support level should not be read as a per-profile optimum. Each panel tracks recall AUC as a function of \(\epsilon\) for one perturbation profile. The panels make clear that the support sensitivity differs across profiles: some curves are nearly flat across the useful range, whereas others show a narrower optimum or even a different local maximum. This is why the main paper treats \(\epsilon=10^{-10}\) as a conservative common operating point rather than as a claim of universal optimality.

\subsection{Figure S8: Descriptive sweep-level bubble overview}
\begin{figure}[t]
\centering
\includegraphics[width=0.95\linewidth]{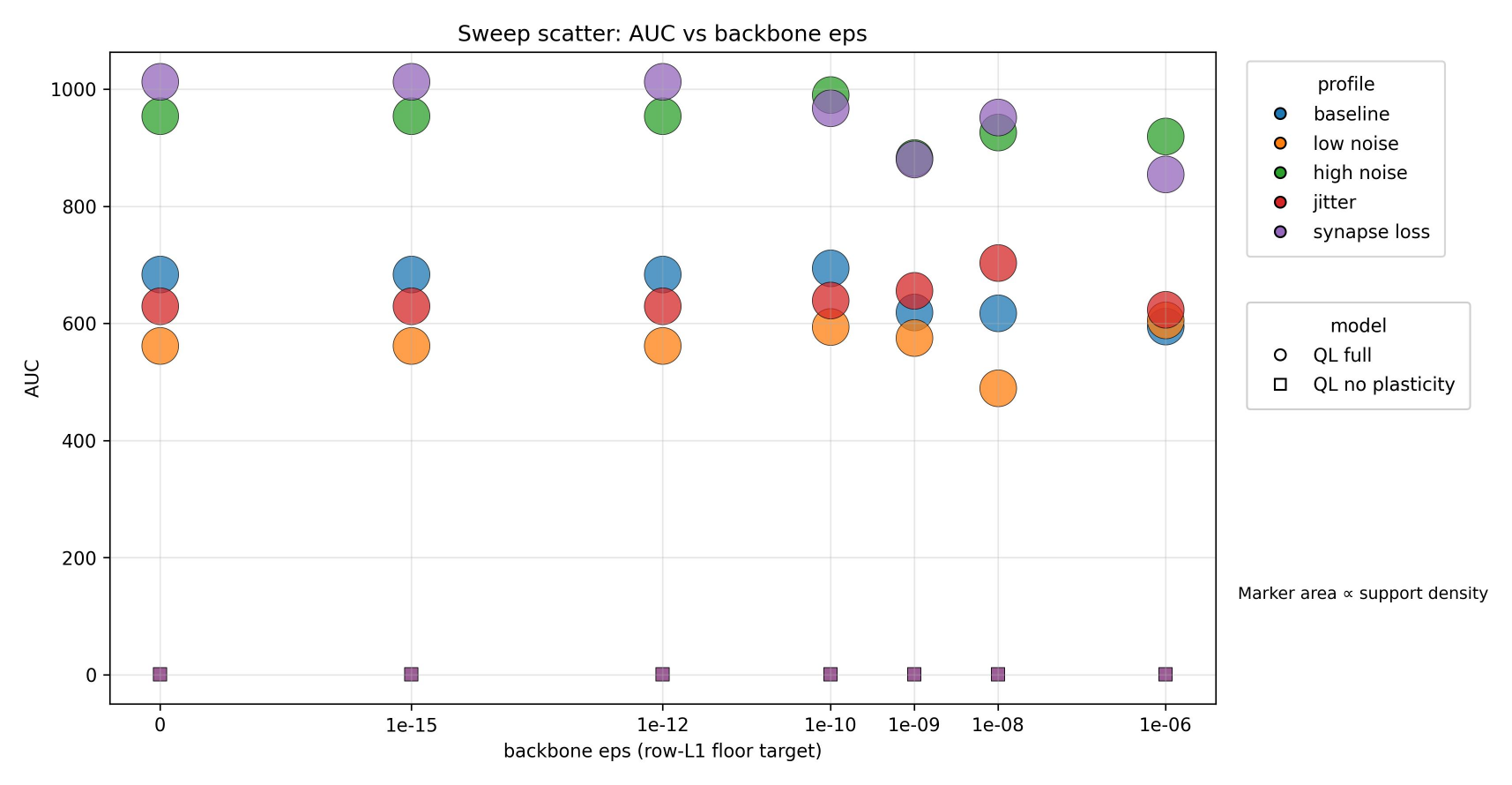}
\caption{Descriptive sweep-level summary of recall AUC and support density across model families and perturbation profiles. Marker size encodes support density, color encodes profile, and shape encodes model family.}
\label{fig:supp_sweep_bubble}
\end{figure}

Figure~\ref{fig:supp_sweep_bubble} provides a descriptive sweep-level overview. It overlays support density, recall AUC, perturbation profile, and model family in one sweep-level view. Because it combines several encoded quantities in a single view, it is best read as an orienting summary of where the full model, the no-plasticity ablation, and the classical controls sit across the sweep.

\subsection{Figure S9: Cross-family backbone-effect summary}

\begin{figure}[t]
\centering
\includegraphics[width=0.99\linewidth]{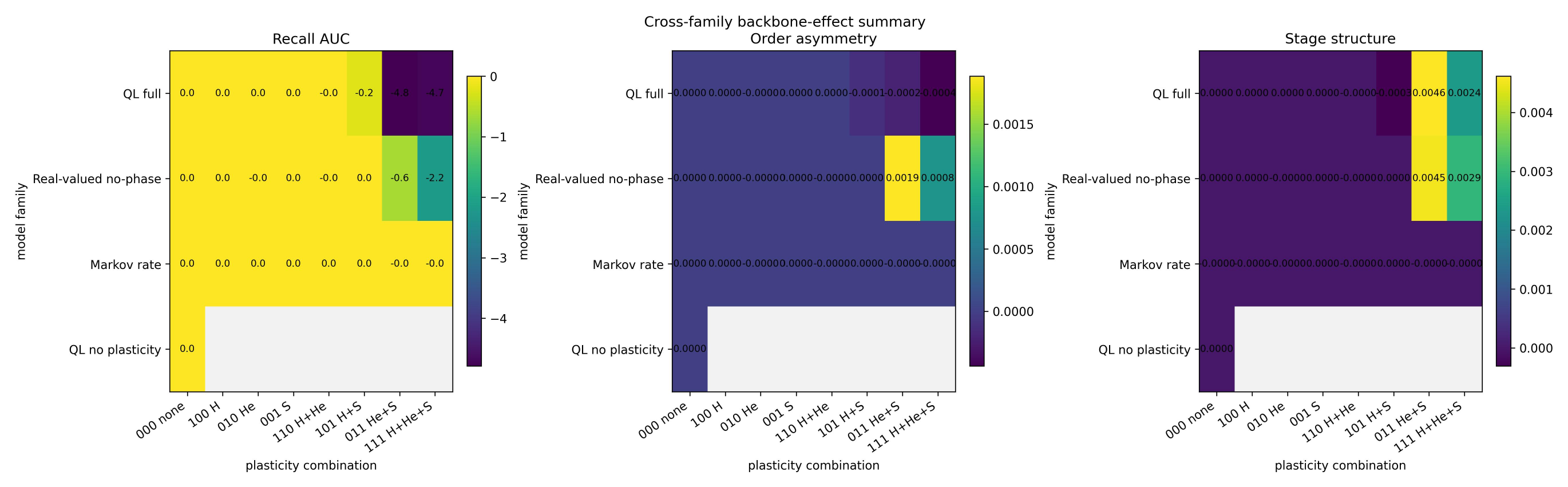}
\caption{Profile-averaged backbone-on minus backbone-off deltas across model family and plasticity combination, shown separately for recall AUC, order asymmetry, and stage structure. The near-zero classical-control panels show that the backbone effect is not a generic property of the factorial grid and is best interpreted as a weak operating constraint on the quantum-like family.}
\label{fig:supp_backbone_delta_family}
\end{figure}

Figure~\ref{fig:supp_backbone_delta_family} summarizes the cross-family backbone effects. When the weak floor is toggled on and off inside the frozen factorial benchmark, the Markov-rate family remains essentially flat across recall AUC, order asymmetry, and stage structure, and the real-valued no-phase family moves only modestly. The quantum-like family, by contrast, retains small but nonzero support sensitivity, especially in recall AUC. This asymmetry is methodologically important because it shows that the selected floor is not acting as a generic booster applied to the whole benchmark. Instead, it behaves as a weak operating constraint whose downstream consequences remain concentrated in the family for which support selection was originally performed.

\subsection{Figure S10: Ablation gap under the no-extra-plasticity condition}
\begin{figure}[t]
\centering
\includegraphics[width=0.82\linewidth]{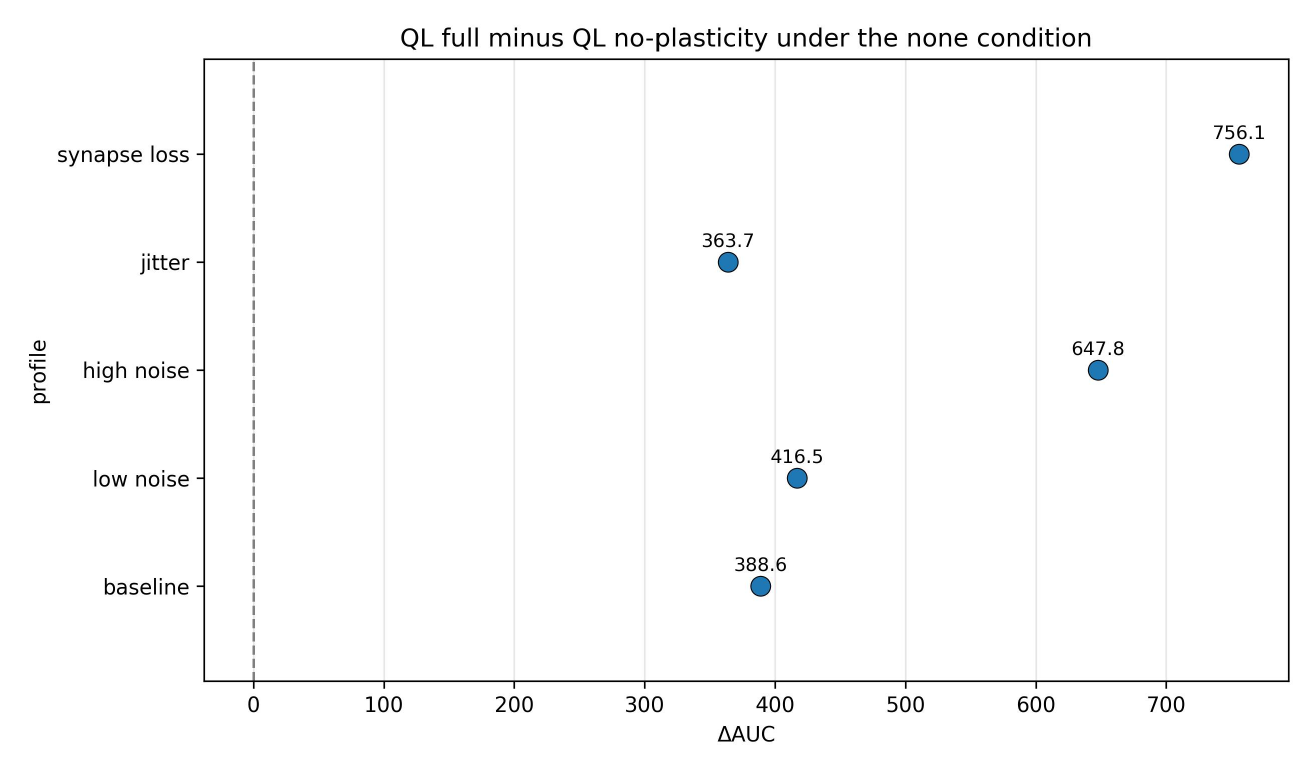}
\caption{Recall AUC gap between the full quantum-like model and the no-plasticity ablation under the no-extra-plasticity factorial condition, shown separately for each perturbation profile.}
\label{fig:supp_ablation_gap_none}
\end{figure}

Figure~\ref{fig:supp_ablation_gap_none} focuses on the most conservative factorial slice: the no-extra-plasticity condition. Even there, the full quantum-like model retains a large recall advantage over the no-plasticity ablation, with the largest gaps again appearing in high noise and synapse loss. This shows that the selected-support recall advantage is already present before the more complex plasticity combinations are layered on top.

Figures~S11--S12 summarize the representative recall--order trade-off and the best-AUC model-placement view used to interpret the cross-family comparison.

\subsection{Figure S11: Representative-point trade-off summary}

\begin{figure}[t]
\centering
\includegraphics[width=0.92\linewidth]{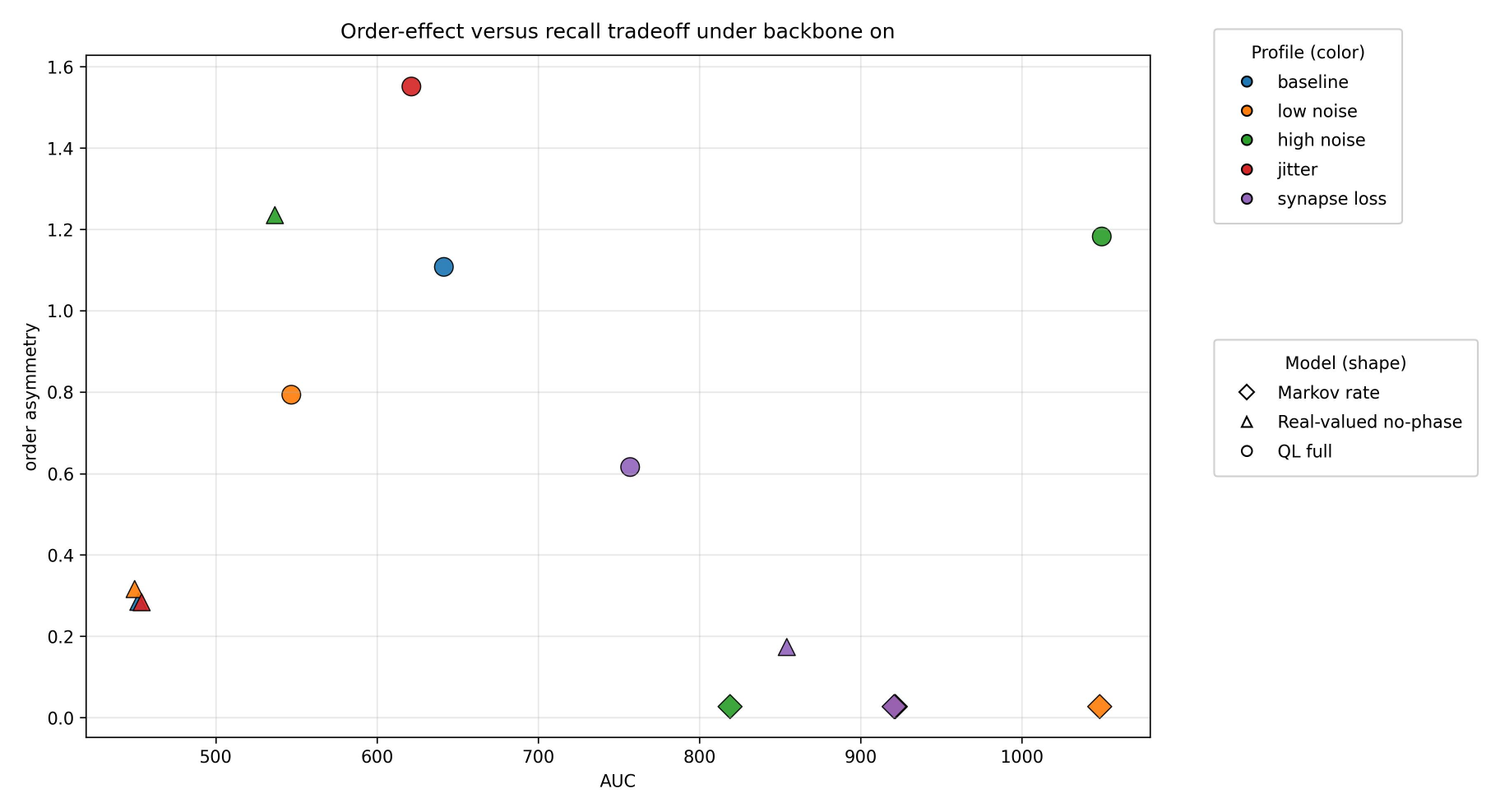}
\caption{Compact backbone-on trade-off summary using the best-AUC representative point for each model--profile pair, plotted as recall AUC versus order asymmetry. For each model family and perturbation profile, the plotted point is the single backbone-on evaluation cell with maximal recall AUC. Marker shape denotes model family and marker color denotes perturbation profile. Some representative points are nearly coincident and therefore overlap visually, most notably for the real-valued no-phase baseline and jitter representatives and for the Markov-rate baseline, jitter, and synapse-loss representatives; the exact selected cells and coordinates are listed in Table~\ref{tab:best_auc_tradeoff_points}.}
\label{fig:supp_order_auc_tradeoff}
\end{figure}

Figure~\ref{fig:supp_order_auc_tradeoff} provides a compact backbone-on trade-off summary in which each model--profile pair is reduced to its single best-AUC representative cell. Relative to the full evaluation-cloud view, this compressed representation makes the family-level separation easier to read while preserving the central recall--order comparison. The exact representative cells and their plotted coordinates are listed in Table~\ref{tab:best_auc_tradeoff_points}.

This reduction yields 15 plotted points in total: five for the quantum-like model with plasticity, five for the real-valued no-phase control, and five for the Markov-rate control. In this view, the Markov-rate control remains concentrated in the high-recall but near-zero-order-asymmetry region, the real-valued no-phase control occupies an intermediate regime, and the quantum-like model with plasticity occupies the higher-order-asymmetry region while remaining recall-competitive in selected profiles.

Not all 15 representatives are visually distinguishable because some selected points are nearly coincident in the recall-AUC--order-asymmetry plane. The real-valued no-phase baseline and jitter representatives lie very close to one another, and the Markov-rate baseline, jitter, and synapse-loss representatives are tightly clustered at essentially the same order-asymmetry value. Table~\ref{tab:best_auc_tradeoff_points} is therefore provided so that the exact selected coordinates can be read directly even where markers overlap visually.

\subsection{Figure S12: Best-AUC model--profile overview}
\begin{figure}[t]
\centering
\includegraphics[width=0.88\linewidth]{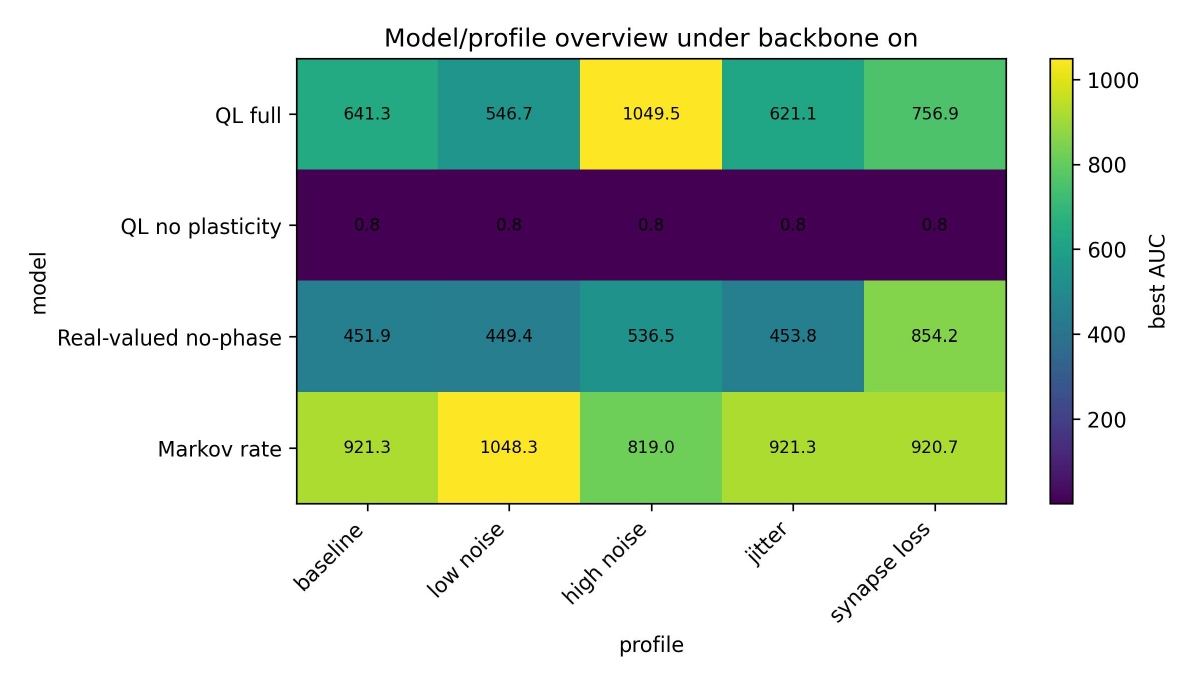}
\caption{Best-AUC profile-by-model overview under backbone-on conditions. Cell values report the representative recall AUC for each model--profile pair. The figure provides descriptive context for the raw recall pattern in isolation; inference in the main text is based instead on the joint recall--order--stage-structure benchmark.}
\label{fig:supp_best_auc_overview}
\end{figure}

Figure~\ref{fig:supp_best_auc_overview} provides a descriptive profile-by-model view of the best-AUC representatives under backbone-on conditions. Its main value is to make the scale separation across model families immediately visible: the Markov-rate control is strongest on raw recall AUC, the real-valued no-phase control is intermediate, and the no-plasticity ablation remains near zero. Because this figure isolates recall alone, it serves as descriptive context for the raw recall separation across model families. The main conclusions rely on the joint recall--order--stage-structure benchmark.

Figures~S13--S15 provide the factorial plasticity summaries behind the main-paper plasticity results.

\subsection{Figure S13: Pairwise interaction heatmap}
\begin{figure}[t]
\centering
\includegraphics[width=0.98\linewidth]{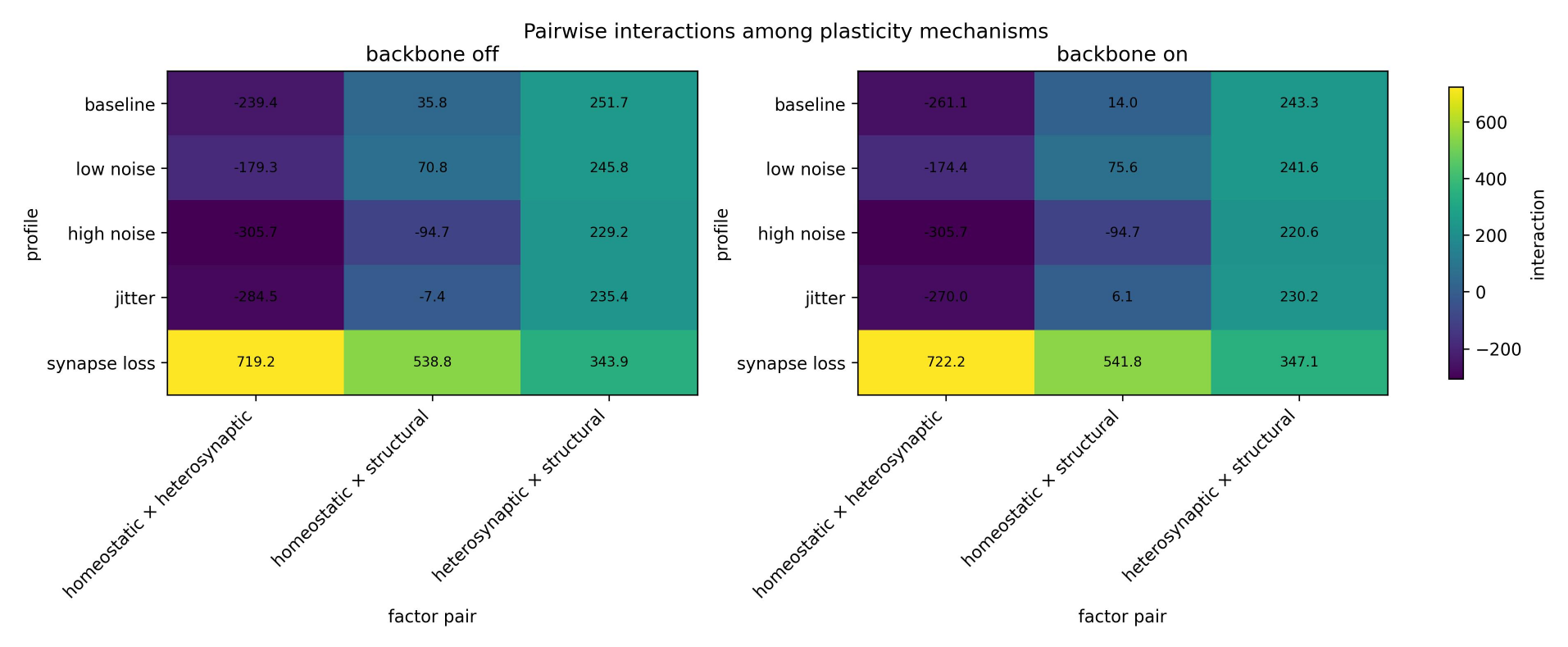}
\caption{Pairwise interaction terms among homeostatic, heterosynaptic, and structural plasticity mechanisms for recall AUC, shown separately for support-off and support-on regimes.}
\label{fig:supp_pairwise}
\end{figure}

Figure~\ref{fig:supp_pairwise} gives the full two-factor interaction map that sits behind the factorial decomposition summarized in Fig.~\ref{fig:main_effects_auc}. The most important pattern is the strong positive heterosynaptic-structural interaction, with a weaker positive homeostatic-structural interaction and a mildly negative homeostatic-heterosynaptic interaction. These maps show that the plasticity effects are not purely additive and that interaction strength depends on both support mode and on perturbation profile.

\subsection{Figure S14: Full factorial recall heatmap}

\begin{figure}[t]
\centering
\includegraphics[width=0.98\linewidth]{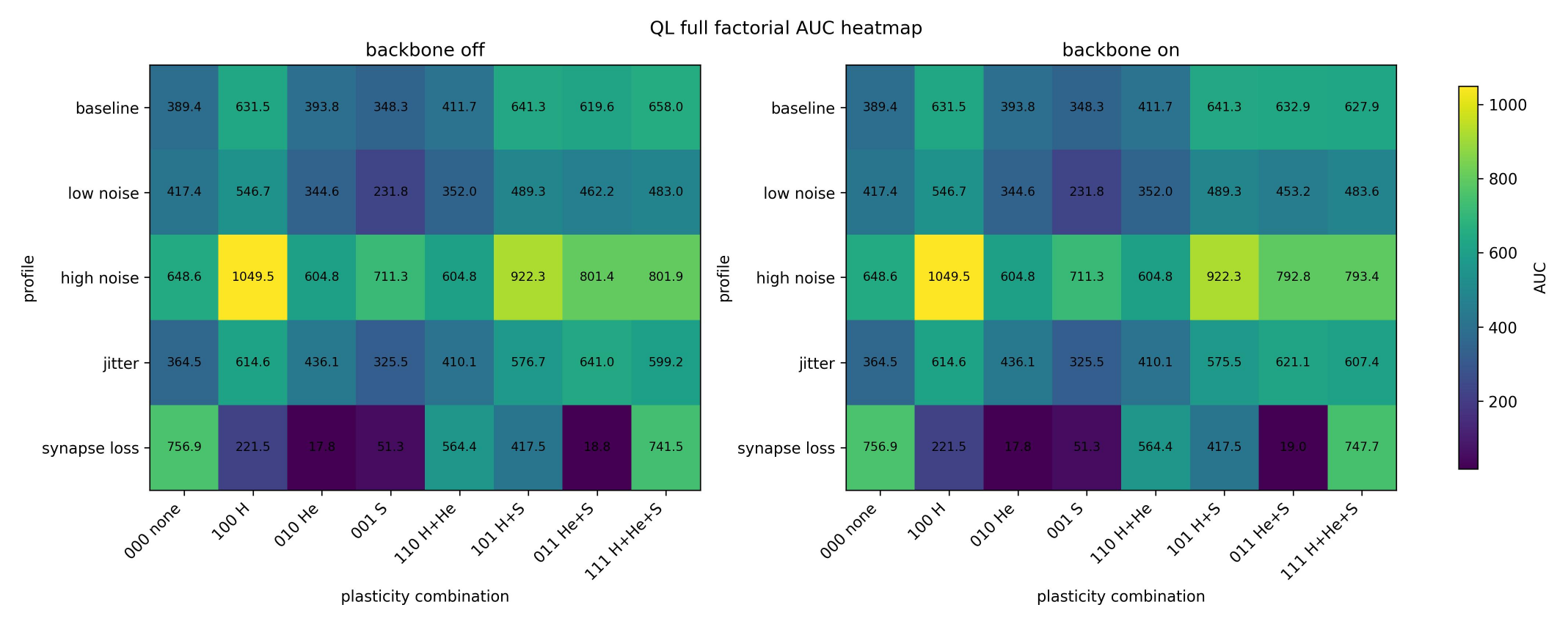}
\caption{Full factorial recall AUC heatmap for the full quantum-like model across perturbation profiles, plasticity combinations, and support conditions.}
\label{fig:supp_factorial_full}
\end{figure}

Figure~\ref{fig:supp_factorial_full} is the complete factorial AUC map underlying the compressed main-effects summary in Fig.~\ref{fig:main_effects_auc} and the delta-from-none view in Fig.~\ref{fig:supp_factorial_delta_from_none}. It shows the full \(2^3\) cell pattern rather than only the averaged contrasts. Positive recall does not arise uniformly from adding more plasticity; specific combinations are favorable, neutral, or detrimental depending on profile and support condition.

\subsection{Figure S15: Delta-from-none factorial recall heatmap}

\begin{figure}[t]
\centering
\includegraphics[width=0.98\linewidth]{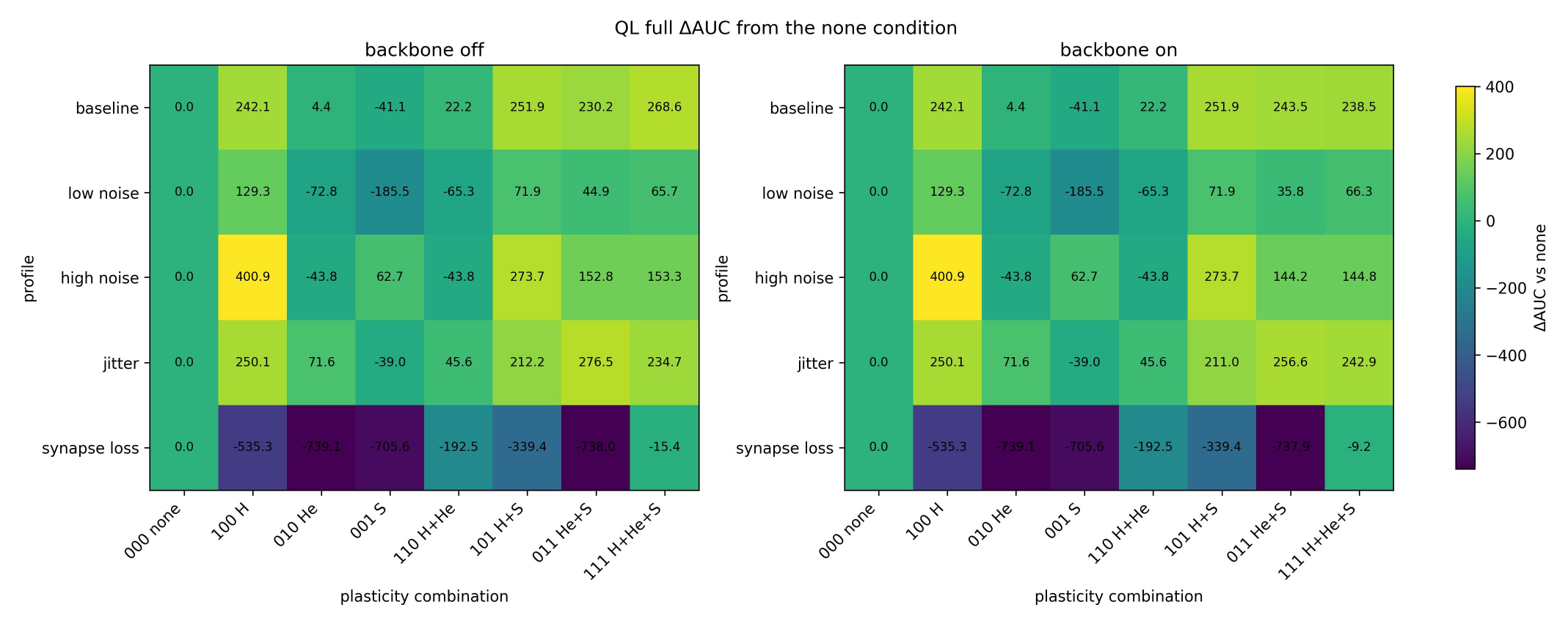}
\caption{Factorial recall change relative to the no-extra-plasticity reference within each support regime. Each cell reports the mean change in $\mathrm{AUC}_{\mathrm{recallA}}$ relative to the all-off plasticity condition for one profile and one plasticity combination. Positive values identify combinations that improve recall beyond the baseline scaffold-only setting, whereas negative values indicate that the added mechanism mixture is detrimental in that context.}
\label{fig:supp_factorial_delta_from_none}
\end{figure}

Figure~\ref{fig:supp_factorial_delta_from_none} asks which plasticity combinations improve recall relative to the no-extra-plasticity reference. The broadest positive regions appear when homeostatic plasticity is present, whereas the largest gains from structural and heterosynaptic plasticity are more patchy and combination-dependent. This panel complements the main-effect summary by showing the full profile-by-combination pattern rather than only the average first-order contrasts.

Figures~S16--S18 provide the broader control-family factorial views for recall, stage structure, and order asymmetry.

\subsection{Figure S16: Model-family factorial recall overview}
\begin{figure}[t]
\centering
\includegraphics[width=0.98\linewidth]{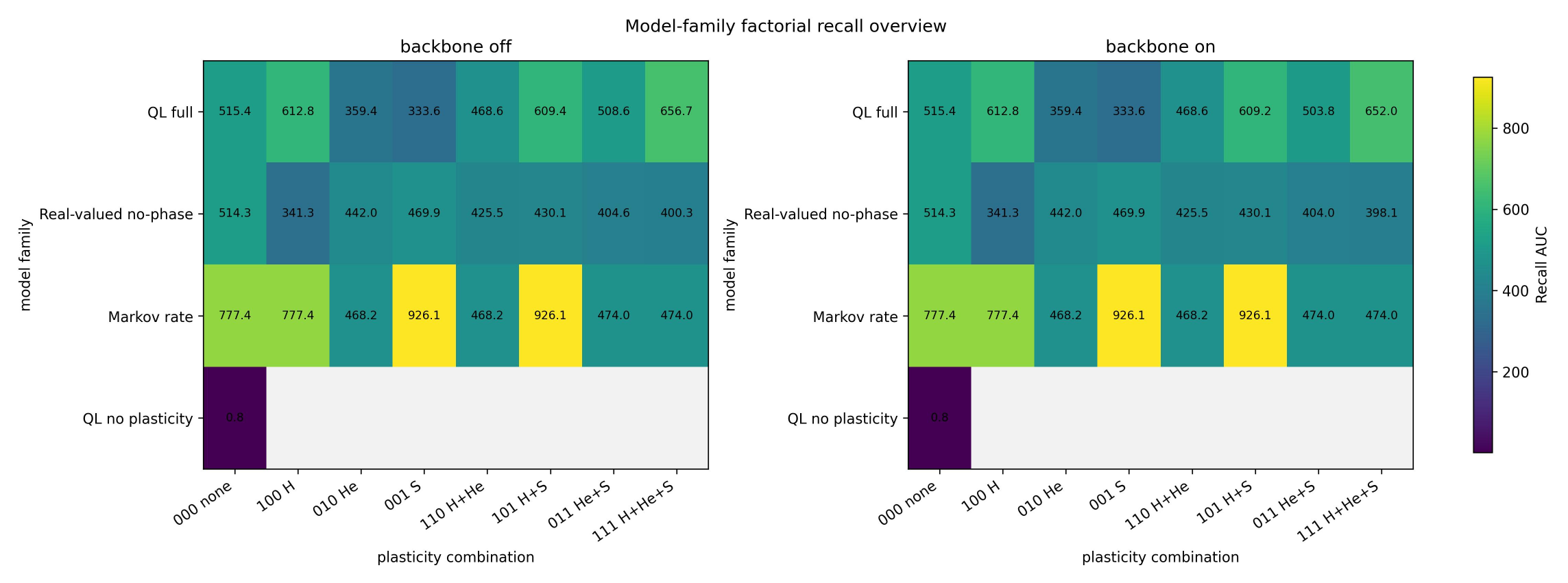}
\caption{Profile-averaged factorial recall AUC across model family, plasticity combination, and backbone mode. Rows give model family, columns give plasticity combination, and the two panels separate support-off from support-on conditions. Blank cells mark combinations that were not executed, which occurs only for the quantum-like no-plasticity ablation.}
\label{fig:supp_model_family_auc_overview}
\end{figure}

Figure~\ref{fig:supp_model_family_auc_overview} compares the model families on the same profile-averaged plasticity grid for recall AUC. The Markov-rate family responds strongly along the scalar recall axis, the real-valued control is more mixed, and the quantum-like family reaches its strongest recall under all-three or homeostatic-containing combinations.

\subsection{Figure S17: Model-family factorial stage structure overview}
\begin{figure}[t]
\centering
\includegraphics[width=0.98\linewidth]{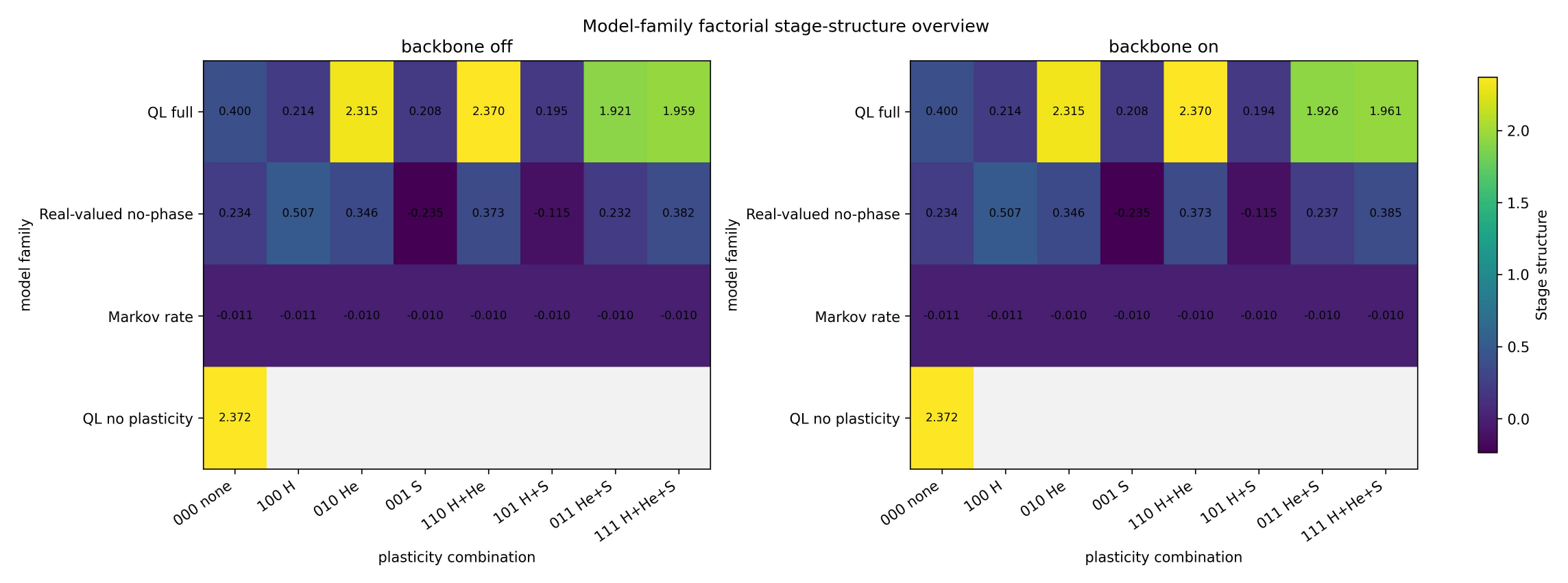}
\caption{Profile-averaged factorial stage-structure score across model family, plasticity combination, and backbone mode. The layout matches Fig.~\ref{fig:supp_model_family_auc_overview} so that the control-family differences can be compared metric by metric.}
\label{fig:supp_model_family_stage_overview}
\end{figure}

Figure~\ref{fig:supp_model_family_stage_overview} shows the same factorial grid through the stage-structure score. The Markov-rate control remains near zero or slightly negative. The real-valued no-phase control shows some positive stage structure but does not reach the same range as the quantum-like family, which retains the strongest positive stage-structure values for heterosynaptic-containing combinations. 

\subsection{Figure S18: Model-family factorial order-asymmetry overview}

\begin{figure}[t]
\centering
\includegraphics[width=0.98\linewidth]{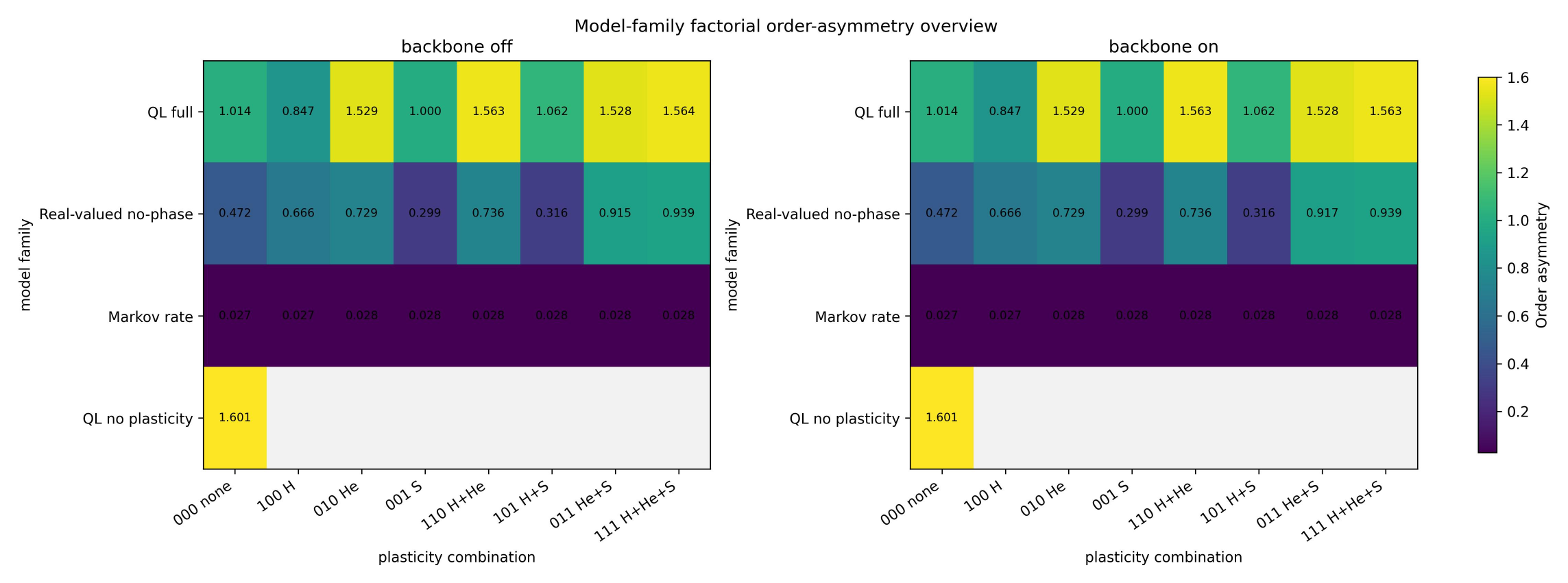}
\caption{Profile-averaged factorial order asymmetry across model family, plasticity combination, and backbone mode. The Markov-rate control remains low-order across the grid, the real-valued no-phase control is intermediate, and the quantum-like family remains highest on this benchmark-specific order-sensitive summary.}
\label{fig:supp_model_family_order_overview}
\end{figure}

Figure~\ref{fig:supp_model_family_order_overview} compares order-asymmetry across the shared factorial grid. The Markov-rate family remains confined to a narrow low-order band regardless of plasticity combination, the real-valued no-phase family reaches an intermediate regime, and the full quantum-like family occupies the highest-order region. Thus, even when the classical families traverse the same plasticity grid, they do not recreate the same order-sensitive operating region.

\end{document}